\documentclass[aps,10pt,prd,groupedaddress,nofootinbib,notitlepage,eqsecnum,preprintnumbers]{revtex4-2}

\usepackage[utf8]{inputenc}
\usepackage{amsmath}
\usepackage{amssymb}
\usepackage{graphicx}
\usepackage{dcolumn}   
\usepackage{bm}        
\usepackage{amssymb}   
\usepackage{subcaption}
\usepackage{mathtools}
\usepackage{xcolor}
\usepackage{hyperref}
\usepackage{ulem}

\begin{document}

\newcommand{\Red}[1]{{\color{red}#1}}
\newcommand{\mb}{\mathbf}
\newcommand{\mpl}{M_{\mathrm{Pl}}}
\newcommand{\GW}{\mathrm{GW}}
\newcommand{\rh}{\mathrm{rh}}
\newcommand{\cH}{\mathcal{H}}
\newcommand{\calH}{{\cal H}}
\newcommand{\calL}{{\cal L}}
\newcommand{\calO}{{\cal O}}
\newcommand{\AR}{\mathcal{A}_\mathcal{R}}
\newcommand{\calP}{{\cal P}}
\newcommand{\calR}{{\cal R}}
\newcommand{\be}{\begin{equation}}
\newcommand{\ee}{\end{equation}}
\newcommand{\bea}{\begin{eqnarray}}
\newcommand{\eea}{\end{eqnarray}}
\newcommand{\nn}{\nonumber}
\newcommand{\bx}{\mathbf{x}}
\newcommand{\bk}{\mathbf{k}}
\newcommand{\bl}{\mathbf{l}}
\newcommand{\bp}{\mathbf{p}}
\newcommand{\bq}{\mathbf{q}}
\newcommand{\vf}{\varphi}
\newcommand{\df}{\delta\vf}
\newcommand{\tq}{{\widetilde Q}}
\newcommand{\tf}{{\widetilde\vf}}
\newcommand{\fb}{{\bar\vf}}
\newcommand{\fnl}{f_\text{NL}}
\newcommand{\dN}{{\text{d}n}}
\newcommand*{\dif}{\mathop{}\!\mathrm{d}}
\newcommand{\ImPart}{\mathrm{Im}}
\newcommand{\AW}[1]{{\color{blue}(AW: #1)}}
\newcommand{\AWadd}[1]{{\color{blue}#1}}

\preprint{YITP-24-39}

\title{Revisiting the Ultraviolet Tail of the Primordial Gravitational Wave}
	\author{Shi Pi${}^{a,b,c}$} \email{shi.pi@itp.ac.cn} 
	\author{Misao Sasaki${}^{c,d,e}$}\email{misao.sasaki@ipmu.jp}
	\author{Ao Wang${}^{a,f}$}\email{wangao@itp.ac.cn}
	\author{Jianing Wang${}^{a,f}$}\email{wangjianing@itp.ac.cn}·
		\affiliation{
		$^{a}$ CAS Key Laboratory of Theoretical Physics, Institute of Theoretical Physics, Chinese Academy of Sciences, Beijing 100190, China \\
            $^{b}$ Center for High Energy Physics, Peking University, Beijing 100871, China \\
		$^{c}$ Kavli Institute for the Physics and Mathematics of the Universe (WPI), The University of Tokyo, Kashiwa, Chiba 277-8583, Japan\\
		$^{d}$ Center for Gravitational Physics and Quantum Information,
  Yukawa Institute for Theoretical Physics, Kyoto University, Kyoto 606-8502, Japan\\
		$^{e}$ Leung Center for Cosmology and Particle Astrophysics,\\
  National Taiwan University, Taipei 10617,\\
  $^{f}$ School of Physical Sciences, University of Chinese Academy of Sciences, Beijing 100049, China}
	\date{\today}

\begin{abstract}
High-frequency primordial gravitational waves (PGWs) with wave numbers larger than the Hubble parameter at the end of inflation are originated from the ultraviolet (UV) modes, which are never stretched out of the horizon. Such a UV tail of the PGW energy spectrum has a spurious logarithmic divergence. We study the origin of such a divergence, and find that it comes from the instantaneous inflation-to-post-inflation transition, which can be removed by considering a finite duration. For the first time, we obtain a semi-analytical expression for the PGW energy spectrum. We find that the UV tail decays exponentially, while the decay rate depends solely on the transition rate. When there is a stiff post-inflationary stage, the enhanced PGW displays a characteristic spectral shape of power-law increasing and exponential decaying. We propose a fitting formula which can be used for signal searching.

\end{abstract}

\maketitle

\section{Introduction}

Inflation can solve the horizon, flatness, and monopole problems of the hot big bang cosmology \cite{Brout:1977ix, Guth:1980zm, Starobinsky:1980te, Linde:1981mu, Albrecht:1982wi}. During inflation, the quantum fluctuations are generated on very small scales, then stretched out of the Hubble horizon and frozen by the expansion of the inflationary universe \cite{Mukhanov:1981xt}. After reentering the Hubble horizon in the decelerated expansion stage, the primordial scalar perturbation serves as the seeds for the cosmic microwave background (CMB) anisotropies and the large scale structure we observe today, which is now accurately measured to be of order $10^{-5}$ at scales larger than 1 Mpc \cite{Planck:2018jri}. The primordial tensor perturbations reentering the Hubble horizon will turn to be the primordial gravitational waves (PGWs), which are believed to be the smoking gun of the inflationary paradigm. There are many ongoing or planned experiments, for instance, BICEP/Keck Array\cite{BICEP:2021xfz}, NANOGrav \cite{Arzoumanian_2020}, EPTA \cite{EPTA:2015qep}, PPTA \cite{Zhu:2014rta}, CPTA \cite{Xu:2023wog}, LISA \cite{Barausse:2020rsu, LISA:2022kgy}, Taiji \cite{Hu:2017mde, Ruan:2018tsw} and TianQin \cite{TianQin:2015yph}, aiming to probe PGW in vast frequency ranges. However, it is difficult to detect such waves in the cosmological concordance model, because the standard cosmology has a radiation-dominated era right after the end of inflation, which predicts a flat spectrum of GW energy spectrum $\Omega_\GW(f)\sim10^{-16}(r/0.01)$ spanning from $10^{-16}$ Hz to $10^8$ Hz \cite{Boyle:2005se,Kuroyanagi:2008ye,Watanabe:2006qe,Saikawa:2018rcs}, with a negligible tensor tilt $n_T\approx0$ , where $r=\mathcal{P_R}/\mathcal{P}_T\lesssim0.01$ is the tensor-to-scalar ratio. 

Stretching the fluctuations out of the Hubble horizon is equivalent to the classicalization of the quantum fluctuations, which is crucial for PGWs to be observable. The flat energy spectrum of the PGW extends to such a high frequency $f_\mathrm{UV}$ where the primordial tensor perturbation is never stretched out of the horizon. Above such a high frequency, a simple estimation gives a quartic ultraviolet (UV) divergence $\Omega_\GW\propto f^4$ \cite{Kuroyanagi:2008ye}, which has to be removed. Similar to the UV divergence of the energy-momentum tensor or two-point function of a scalar field in the curved space, a UV regularization can be used to remove the divergence order by order until the remaining regularized quantity is proportional to $f^{-2}$ \cite{Parker:1974qw, Fulling:1974pu}. It was found that the adiabatic regularization can affect the infrared spectrum, thus leading to ambiguous interpretations \cite{Durrer:2009ii, Marozzi:2011da, Markkanen:2017rvi}. The adiabatic regularization is applied to the PGW energy spectrum in \cite{Wang:2015zfa, Zhang:2018dvc}. 

For the short-wavelength primordial tensor perturbations which are never stretched out of the horizon, relic gravitons are generated out of the vacuum as the vacuum state changes after inflation. In such a case, the vacuum fluctuations should be removed appropriately as they only contribute to the background but not to the energy density $\Omega_\GW$ defined by short-wave expansion \cite{Isaacson:1968hbi}. Only the excited relic gravitons can contribute to the energy spectrum. Such considerations are firstly done in \cite{Grishchuk:1974ny,Ford:1977dj} and reviewed in \cite{Birrell:1982ix,Parker:2009uva}. It was found that the number density of the created gravitons has a Boltzmann suppression, which implies $\Omega_\GW\propto\exp(-\mu k/k_*)$, where $\mu$ depends on the transition time to the post-inflation era. In the instantaneous reheating case $\mu=0$, the Boltzmann suppression factor is absent, leaving a flat PGW spectrum that suffers logarithmic divergence \cite{Wang:2015zfa}. However, numerical simulations show that the typical duration of the transition should be $H^{-1}$ \cite{Lozanov:2016hid,Lozanov:2017hjm,Antusch:2020iyq,Saha:2020bis}, which leads to a non-zero decay rate $\mu$. Therefore, the exponential suppression makes the UV part convergent, and no regularization is needed. The decay rate $\mu$ was only calculated numerically \cite{Giovannini:2008tm}, which must depend on the transition rate of the inflationary stage to the radiation (or $w$-dominated) stage. One may concern about the physical interpretation of this decay rate, and how it depends on the transition rate of the background evolution. This is the main topic of this paper. 

In this paper, we parameterize $\calH(\eta)$ by an analytical function which reflects the smooth transition from inflation to the post-inflationary era and calculate the decay rate of the PGW spectrum by solving the equation of motion of the tensor perturbation by the Green function method.  We find that as $\calH(\eta)$ connects inflation ($\propto1/\eta$) and deccelerated expansion ($\propto\eta^n$), it always has poles in the complex plane of $\eta$, and the decay rate $\mu$ is proportional to the amplitude of the first pole of $\calH(\eta)$, which is monotonic to the inflation-to-post-inflation transition rate. This conclusion is robust, independent of the concrete functional form of $\calH(\eta)$. While the infrared power-law scaling reflects the thermal history of the universe or the primordial tilt, the decay rate to the UV side of the peak can be used to probe the transition rate of inflation to the radiation- or $w$-dominated era. 

In the standard cosmology, due to the minimal e-folding number required by inflation, such an exponential tail appears at a high frequency of $\sim10^{8}\mathrm{Hz}$, with an amplitude of $\Omega_\GW(f)\sim10^{-16}(r/0.01)$, thus is difficult to probe. However, for some special models, for instance, the non-standard post-inflationary expansion with a stiff equation-of-state parameter $w>1/3$ including the so-called kination-dominated era \cite{Spokoiny:1993kt,Joyce:1996cp,Ferreira:1997hj,Peebles:1998qn,Tashiro:2003qp}), or for a blue-tilted primordial tensor spectrum \cite{Cook:2011hg,Biagetti:2013kwa,Ashoorioon:2013eia,Brandenberger:2014faa,Gong:2014qga,Wang:2014kqa,Ashoorioon:2014nta,Mohanty:2014kwa,Mohanty:2014kwa,Mukohyama:2014gba} (see Ref.~\cite{Mukohyama:2014gba} for a review and references therein), the GW spectrum $\Omega_\GW(f)$ will increase towards the high frequencies, and peak around $f_\mathrm{UV}$ may be detectable in the future \cite{Aggarwal:2020olq}. Earlier discussions on this topic can be found for instance in Refs.\cite{Giovannini:1998bp,Peebles:1998qn,Tashiro:2003qp,Giovannini:2009kg,Liu:2015psa,Figueroa:2018twl,Giovannini:2019oii,Bernal:2019lpc,Figueroa:2019paj,DEramo:2019tit,Gouttenoire:2021jhk}. After the recent announcement of the detection of a nanohertz stochastic GW background by the Pulsar Timing Array (PTA) collaborations NANOGrav~\cite{NANOGrav:2023gor,NANOGrav:2023hde}, EPTA combined with InPTA~\cite{EPTA:2023fyk,EPTA:2023sfo,EPTA:2023xxk}, PPTA\,\cite{Zic:2023gta,Reardon:2023gzh,Reardon:2023zen}, CPTA\,\cite{Xu:2023wog}, as well as IPTA \cite{InternationalPulsarTimingArray:2023mzf}, the interpretation in terms of PGWs from a stiff post-inflation stage or the blue-tilted primordial tensor spectrum has attracted much attention \cite{NANOGrav:2023hvm,Vagnozzi:2020gtf,Bhattacharya:2020lhc,Kuroyanagi:2020sfw,Li:2021htg,Benetti:2021uea,Vagnozzi:2023lwo,Fu:2023aab,Borah:2023sbc,Datta:2023vbs,Figueroa:2023zhu,Choudhury:2023kam,Ben-Dayan:2023lwd,Jiang:2023gfe,Ellis:2023oxs,Oikonomou:2023bli,Datta:2023xpr,Harigaya:2023pmw,Choi:2023tun,Ye:2023tpz}. 
In our work, we further find that the start point of the exponential decay of the PGW spectrum, originating from adiabatic vacuum fluctuations of the tensor perturbation, at higher frequencies must lie on this line: $\Omega_{\text{GW}}(f_{\text{UV}}) \sim 10^{-16}\left(f_{\text{UV}}/10^8\text{Hz}\right)^4$, where $f_{\text{UV}}$ is the minimum frequency that never exits the horizon. Moreover, we find that fine-tuning is required for the exponential decay feature of a purely blue-tilted PGW spectrum to explain the PTA data appearing in the PTA band or nearby. Otherwise, it will be in conflict with the CMB constraint on relativistic components unless a broken power-law spectrum is introduced\,\cite{Jiang:2023gfe,Oikonomou:2023bli,Ye:2023tpz}.

The paper is organized as follows. In Sec.\ref{sec:primordial_gravitational_wave}, we review how the primordial gravitational waves are produced during cosmic expansion. 
In Sec.\ref{sec:smoothing_transition_model}, the Born approximation results and the numerical results in the cases of the inflation-radiation model and inflation-stiff equation of state (EoS) model are demonstrated. 
In Sec.\ref{Fitting formula}, we discuss a fitting formula describing the whole spectrum around the peak.
Sec.\ref{sec:conclusion} gives the conclusion of our paper. 

\section{Primordial gravitational wave} 
\label{sec:primordial_gravitational_wave}
In cosmological perturbation theory, up to second-order perturbation in the Lagrangian, the scalar mode and the tensor mode are decoupled, known as the decomposition theorem. Therefore to study the primordial tensor perturbation and PGWs, we can focus on the tensor-type perturbation of the metric. 
In this section, we review the canonical quantization of the tensor perturbation and calculate the energy density spectrum of PGWs we observe today.

\subsection{Action for the tensor perturbation}
\label{sub:action_of_tensor_perturbation}

We assume a spatially-flat homogeneous and isotropic background,
\begin{equation}
    ds^2=\bar{g}_{\mu\nu}dx^\mu dx^\nu=a^2(\eta)f_{\mu\nu}dx^\mu dx^\nu\,,
\end{equation}
where $\eta$ is the conformal time and $\eta_{\mu\nu}$ is the flat Minkowski metric. On this background, we consider a perturbation,
\begin{equation}
	g_{\mu\nu}=\bar{g}_{\mu\nu}+\delta g_{\mu\nu}.
\end{equation}
To study GWs, we focus on the tensor perturbation, characterized by the transverse-traceless components of the spatial metric,
\begin{eqnarray}
    \delta g_{\mu\nu}=a^2h_{\mu\nu}\,;\quad
     h_{0\mu}=h^{i}{}_{i}=\partial_i h^{ij}=0\,,
\end{eqnarray}
where the indices are raised and lowered by the flat metric.
 The action of the tensor perturbation $h_{ij}$ is
\begin{equation}\label{def:St}
\mathcal{S}=\frac{M_{\rm Pl}^2}{8}\int d^4x\,a^2
\left(
-f^{\mu\nu}\partial_{\mu}h_{ij}\partial_{\nu}h^{ij}
\right),
\end{equation}
where  $M_{\rm Pl}=(8\pi G)^{-1/2}$ is the reduced Planck mass.

To quantize $h_{ij}$, it is convenient to define a canonical field $t_{ij}$, such that
\begin{equation}\label{def:Scan}
    \mathcal{S}=\frac{1}{2}\int d^4x\,a^2\sqrt{-\det|f_{\mu\nu}|}\left[
-f^{\mu\nu}\partial_\mu t^{ij}\partial_\nu t_{ij}\right]\,;\quad 
t_{ij}\equiv \frac{M_{\rm Pl}}{2}h_{ij}\,,
\end{equation}
where we insert an unity $\sqrt{-\det|f_{\mu\nu}|}$ to facilitate the definition of the effective energy-momentum tensor in the following.

The Hamiltonian corresponding to \eqref{def:Scan} is given by
\begin{align}\label{Hamiltonian}
H(\eta)
=\frac{1}{2}\int d^3\mb{x}\, \Bigg[\frac{1}{a^2}\pi^{ij}\pi_{ij}-a^2t^{ij}
\nabla^2t_{ij}\Bigg],
\end{align}
where $\pi^{ij}$ is the conjugated momentum defined by 
$\pi^{ij}(\mb{x},\eta)\equiv
\partial{\mathcal{L}}/\partial{t_{ij}'(\mb{x},\eta)}=a^2t^{ij}{}'(\mb{x},\eta)$. From now on, a prime denotes a derivative with respect to the conformal time. The field $t_{ij}$ is quantized by imposing the equal-time canonical commutation relation,
\begin{eqnarray}\label{eqtccr}
    [t_{ij}(\mb{x},\eta),\,\pi^{k\ell}(\mb{y},\eta)]=i\delta^{(3)}(\mb{x}-\mb{y})
    \delta_{(i}^k\delta_{j)}^\ell\,.
\end{eqnarray}
This can be accomplished by decomposing $t_{ij}$ into Fourier modes, and perform the Fock quantization. Namely, introducing the annihilation and creation operators 
$\hat{a}_{\bk}^\lambda$ and $\hat{a}_\bk^\lambda{}^\dag$, respectively, we have
\begin{equation}\label{vijdecomp}
	t_{ij}(\mb{x},\eta)=\int\frac{d^3\bk}{(2\pi)^{3/2}}\sum_{\lambda}
 \left(\hat{a}_\bk^\lambda t_k(\eta)
 e^{(\lambda)}_{ij}(\bk) e^{i\bk\cdot\mb{x}}
+\hat{a}_{\bk}^\lambda{}^\dag t_k^*(\eta)
e^{(\lambda)}_{ij}{}^*(\bk)e^{-i\bk\cdot\mb{x}} \right),
\end{equation}
where $k=|\bk|$, $\lambda$ represents the two polarizations, and $\hat{a}_{\bk}^\lambda$ and $\hat{a}_\bk^\lambda{}^\dag$ satisfy the canonical commutation relations,
\begin{equation}\label{commutation relation}
    [\hat{a}_{\bk}^{\lambda_1},\,\hat{a}_{\bk'}^{\lambda_2\dag}]
    =\delta^{(3)}(\bk-\bk')\delta_{\lambda_1,\lambda_2}\,, \qquad \text{others}=0.
\end{equation}
The equation of motion for the mode function $t_p$ can be readily derived from
the Hamiltonian \eqref{Hamiltonian},
\begin{eqnarray}\label{eomvp}
t_k''+2\calH t_k'+k^2 t_k=0\,; \quad \calH\equiv\frac{a'}{a}\,.
\end{eqnarray}
The two polarization tensors satisfy
$\bk^ie^{\lambda}_{ij}(\bk)=0$, 
$e^{\lambda}(\bk)^i{}_i=0$ and 
$e^{\lambda_1,ij}(\bk)e^{\lambda_2}_{ij}(\bk)^*=\delta_{\lambda_1,\lambda_2}$.
In the standard $+$ and $\times$
representation the polarization tensors are real, and are given by
\begin{eqnarray}
   e^{+}_{ij}(\bk)=\frac{1}{\sqrt{2}}(\hat{\mb{m}}_i\hat{\mb{m}}_j-\hat{\mb{n}}_i\hat{\mb{n}}_j)\,,
   \nonumber\\
   e^{\times}_{ij}(\bk)=\frac{1}{\sqrt{2}}(\hat{\mb{m}}_i\hat{\mb{n}}_j+\hat{\mb{n}}_i\hat{\mb{m}}_j)\,,
\end{eqnarray}
where $\hat{\mb{m}}$ and $\hat{\mb{n}}$ are two orthogonal unit vectors on the plane perpendicular to the wave vector $\bk$. The equal-time canonical commutation relation is recovered by combining \eqref{commutation relation} and the Klein-Gordon (KG) normalization condition on the mode function $t_k$,
\begin{equation}\label{normalization condition}
    t_k{t'_k}^*-t_k'{t_k}^*=\frac{i}{a^2}\,.
\end{equation}
In the above, we have used the fact that general relativity is parity invariant and that the spacetime is spatially flat. 
The fact that the equation of motion \eqref{eomvp} and the KG normalization condition \eqref{normalization condition} do not uniquely determine the solution for $t_p$ is related to the ambiguity in the choice of the vacuum state. We will specify it later by appealing to a physical argument.

The energy momentum tensor is given by taking the functional derivative of $\mathcal{S}$ with respect to the metric $f^{\mu\nu}$,
\begin{eqnarray}\label{emtensor}
    T_{\mu\nu}=-2\frac{\delta \mathcal{S}}{\delta f^{\mu\nu}}
    =a^2\left[\partial_\mu t^{ij}\partial_\nu t_{ij}-\frac{1}{2}f_{\mu\nu}
    \left(f^{\alpha\beta}\partial_\alpha t^{ij}\partial_\beta t_{ij}\right)\right]\,,
\end{eqnarray}
where we have set $\sqrt{-\det|f_{\mu\nu}|}=1$ after taking the functional derivative.
In particular, the effective GW energy density is given by
\begin{eqnarray}\label{eq: rhogw}
    \rho_{\rm GW}=\frac{1}{a^4}T_{00}=\frac{1}{2a^2} 
\left[t^{ij}{}'t_{ij}'+\partial^k t^{ij}\partial_k t_{ij}\right],
\end{eqnarray}
where $\partial^k=\delta^{ki}\partial_i$ and
the factor $a^{-4}$ is inserted to make it the physical energy density rather than the
conformal energy density. 
As can be easily seen by comparing this expression with \eqref{Hamiltonian}, the Hamiltonian is nothing but the volume integral of the conformal energy density, $H=\int d^3\bm{x}\,a^4T_{00}=a^4\rho_{\rm GW}V$, where $V$ is the conformal volume of the system which is formally infinite, corresponding to the zero of the delta function in momentum space, $V=(2\pi)^3\delta^{(3)}(\bk)|_{\bk=0}$.

Plugging the quantized field \eqref{vijdecomp} to the Hamiltonian \eqref{Hamiltonian}, we obtain 
\begin{eqnarray}\label{hatH}
\hat{H}(\eta)
 =\sum_{\lambda}\int d^3\mb{k}H_\bk^\lambda\,;
\,\,
 H_\bk^\lambda=
 \frac{a^2}{2}\left(\hat{a}_\mb{k} \hat{a}_{-\mb{k}} E[t_k,t_k] +\hat{a}_\mb{k}^\dagger \hat{a}_{-\mb{k}}^\dagger E[t_k^*,t_k^*]+\left(\hat{a}^\dagger_{\mb{k}}\hat{a}_\mb{k} +
 \hat{a}_{\mb{k}}\hat{a}_\mb{k}^\dag\right)E[t_k,t_k^*]\right),
\end{eqnarray}
where
\begin{align}\label{Energy variable}
	E[x_{k},y_k]\equiv x_k'y_k'+k^2x_ky_k\,,
\end{align}
and we have omitted the polarization indices $\lambda$ from the expressions for simplicity.
As we do not consider parity violating cases, 
all the results will be independent of the polarization, and hence will be multiplied by two to account for the two degrees of freedom.

If we take the vacuum expectation value of the Hamiltonian, for the vacuum annihilated by $\hat{a}_\bk$, $\hat{a}_\bk|0\rangle=0$, we obtain
\begin{eqnarray}
    \langle0|\hat{H}|0\rangle
    =a^2\int d^3\bk E[t_k,t_k^*]\,\delta^{(3)}(\bm{p})|_{\bm{p}=0}
    =a^2\int \frac{d^3\bk}{(2\pi)^3}E[t_k,t_k^*]\,V\,,
\end{eqnarray}
which implies the energy density,
\begin{eqnarray}\label{rhovac}
\rho_{\rm GW,vac}=\frac{1}{a^2} \int \frac{d^3\bk}{(2\pi)^3}E[t_k,t_k^*]\,.
\end{eqnarray}
We note that this quantity is UV divergent in general, hence needs to be regularized.
A reasonable method of regularization will be discussed in the next subsection, together with the choice of the vacuum.

\subsection{Initial condition and Analytical solution} 
\label{sub:initial_condition_and_anlytical_solution}

While the modes are deep inside the horizon, one can neglect the expansion of the universe. Then it is reasonable to choose the vacuum such that its mode function behaves like that in the flat spacetime. Such a mode function is called the adiabatic positive frequency function. During inflation, since all modes satisfy the condition $k^2\gg \calH^2(\sim \eta^{-2}$) in the limit $\eta\to-\infty$, the natural choice is
\begin{eqnarray}\label{inf-advac}
t_k(\eta)\to \frac{e^{i\phi_k}}{a\sqrt{2k}}e^{-ik\eta}\quad {\rm for}~ k\eta\to-\infty,
\end{eqnarray}
where the coefficient is determined by the KG normalization condition \eqref{normalization condition}, apart from the arbitrariness of the phase factor $e^{i\phi_k}$.
Under the de Sitter approximation for the inflationary spacetime, $a=(-H\eta)^{-1}$,
the exact solution for the adiabatic mode function is given by
\begin{eqnarray}\label{dSmode}
    t_p=\frac{H}{\sqrt{2k^3}}(1+ik\eta)e^{-ik\eta}\,,
\end{eqnarray}
where we have chosen the phase $\phi_k=-\pi/2$ for convenience.

For the stage after inflation when the universe is dominated by a fluid with the equation of state parameter $w=p/\rho=\text{const.}$, its scale factor is expressed as
\be
a(\eta)=a_{i}(\eta-\eta_{i})^{2/(1+3w)},
\ee
 where $a_i$ and $\eta_{i}$ are determined by the condition that it smoothly matches the scale factor during inflation, and we assumed $-1/3<w<1$, i.e., the universe undergoes decelerated expansion.
 Let us call it a $w$-stage. In the $w$-stage, we can solve \eqref{eomvp} to obtain
\begin{eqnarray}\label{hankel}
t_{k}=\alpha_k u_k+\beta_k u_k^*\,;
  \quad
  u_k=\frac{1}{a(\eta)}\sqrt{\frac{\pi}{4}\left|\eta-\eta_{i}\right|}\,H^{(2)}_{\nu}\left(k\left|\eta-\eta_{i}\right|\right)\,,
  \end{eqnarray}
where $H^{(2)}_{\nu}$ is the Hankel function of the second kind with the order $\nu$ given by
\be\label{n-def}
\nu=\frac{3(1-w)}{2(1+3w)}\,.
\ee

As usual, the Bogoliubov coefficients $\alpha_k$ and $\beta_k$ satisfy the normalization condition,
\be\label{bogoliubov2}
\left|\alpha_k\right|^2-\left|\beta_k\right|^2=1\,,
\ee
and they are determined by evolving the mode function $t_k$ into the $w$-stage. The mode function $u_k$ is chosen so that it gives the natural vacuum state (i.e., the adiabatic positive frequency function in the limit $k\gg\calH$) in the $w$-stage. Namely, we have
\begin{eqnarray}\label{w-admod}
    u_p\to\frac{e^{i\phi_k}}{a\sqrt{2k}}e^{-ik\eta}\quad {\rm for}~k|\eta-\eta_i|=\frac{2}{1+3w}\frac{k}{\calH}\gg1\,.
\end{eqnarray}
In other words, $u_p$ has exactly the same behavior as $t_p$ in the limit $k/\calH\gg1$. This means $\beta_k\to0$ for modes $k\gg\calH_*$,
where $\calH_*$ is the conformal Hubble parameter around the end of inflation. More precisely, as discussed in the next section, $\beta_k$ vanishes exponentially in the limit $k/\calH_*\to\infty$ for any realistic transitions from inflation to the $w$-stage.

Let $\hat{b}_\bk$ be the corresponding annihilation operator that annihilates the $w$-stage vacuum, $\hat{b}_\bk|w\rangle=0$. Then we have
\begin{eqnarray}
 \hat{a}_\bk=\alpha_k^* \hat{b}_\bk+\beta_k^* \hat{b}_{-\bk}^\dag\,,
 \quad
 \hat{a}_\bk^\dag=\alpha_k \hat{b}_\bk^\dag+\beta_k \hat{b}_{-\bk}\,.
\end{eqnarray}
This implies that the expectation value of the number operator $\hat{b}^\dag_\bk\hat{b}_\bk$ with respect to the original vacuum is non-vanishing,
and is given by $\langle0|\hat{b}^\dag_\bk\hat{b}_\bk|0\rangle=|\beta_k|^2$.
Noting that the field $t_{ij}$ can be equivalently expanded in terms of $u_k$ and $\hat{b}_\bk$ (and their complex and Hermitian conjugates, respectively)
in exactly the same form as \eqref{vijdecomp}. In particular, we have exactly the same form for the Hamiltonian \eqref{Hamiltonian} 
by replacing $t_p$ by $u_p$ and $\hat{a}_\bk$ by $\hat{b}_\bk$. Then the vacuum expectation value of the Hamiltonian for the $w$-vacuum becomes
\begin{eqnarray}\label{wvacH}
    \langle w|\hat{H}|w\rangle=a^2\int \frac{d^3\bk}{(2\pi)^3} E[u_k,u_k^*]V\,.
\end{eqnarray}

Now to regularize the energy density we use the fact that the adiabatic positive frequency functions $t_k$ and $u_k$ coincides with each other in the limit $k\gg\calH$. At the $w$-stage, the vacuum annihilated by $\hat{b}_\bk$ should not give any finite energy density. In the standard quantum field theory, this is achieved by normal ordering the creation and annihilation operators, which corresponds to subtracting \eqref{wvacH} from \eqref{rhovac}. Thus the regularized expectation value of the Hamiltonian is
\begin{eqnarray}
     \langle 0|\hat{H}|0\rangle_{\rm reg}
     =a^2\int \frac{d^3\bk}{(2\pi)^3}  (E[t_p,t_p^*]-E[u_k,u_k^*])V\,.
\end{eqnarray}
Hence the properly regularized energy density is given by
\begin{eqnarray}\label{regrho}
   &&\rho^{\rm reg}_{\rm GW}=\frac{1}{a^2}\int d^3\bk (E[t_p,t_p^*]-E[u_k,u_k^*])\,;
\nonumber\\
&&\quad E[t_p,t_p^*]-E[u_k,u_k^*]=2|\beta_k|^2E[u_k,u_k^*]+\alpha_k\beta_k^*E[u_k,u_k]+\alpha_k^*\beta_kE[u_k^*,u_k^*]\,.
\end{eqnarray}
As mentioned before, since $\beta_k$ vanishes exponentially in the limit $k\gg\calH_*$, the UV divergence is now fully removed. 
Furthermore, recalling that the effective GW energy density is well defined only when averaged over the time interval $\Delta\eta\gg1/k$, and that such is possible only on scales sufficiently smaller than the Hubble scale, $k\gg\calH$, we see that $u_k$ may be well-approximated by \eqref{w-admod}, and the terms proportional to $E[u_k,u_k]$ and $E[u_k^*,u_k^*]$ in \eqref{regrho} are averaged to zero.
Thus we finally obtain
\begin{eqnarray}\label{regbarrho}
\overline{{\rho}^{\rm reg}_{\rm GW}}
=\frac{2}{a^2}\int \frac{d^3\bk}{(2\pi)^3} \, |\beta_k|^2 \overline{E[u_k,u_k^*]}
=\frac{1}{\pi^2 a^4} \int dk\, k^3|\beta_k|^2\,,
\end{eqnarray}
where the bar indicates the time averaging. This result has a simple interpretation that there are $2n_{\bk}=2|\beta_k|^2$ particles, each of which has a comoving energy of $k$ per momentum space element $4\pi k^2dk$. The factor $1/a^4$ is to convert the conformal energy density to the physical energy density.

Here, let us introduce a useful quantity that describes the GW energy density spectrum per unit logarithmic frequency interval normalized by the critical density,
\begin{equation}\label{Ogw}
\Omega_{\GW}(k,\eta)
\equiv\frac{1}{\rho_c}\frac{d\overline{{\rho}^{\rm reg}_{\GW}}}{d\ln k}
=\frac{ k^4}{3\pi^2\mpl^2 H^2 a^4} |\beta_k|^2.
\end{equation} 
This is the GW spectrum we observe during the $w$-stage. 
In this paper, we are particularly interested in the high-frequency GWs whose wavelengths never exceed the Hubble horizon scale much. 

\section{Smoothing transition Model}
\label{sec:smoothing_transition_model}
The standard model of cosmology assumes that inflation is followed by the radiation-dominated era, subsequently by the matter-dominated era, and finally by the dark-energy-dominated era. 
However,  since the transition from inflation to the radiation era is not well understood yet, as there is virtually no observational evidence, we consider not only the transition to the radiation-dominated stage but also the transition to a $w$-stage with $w>1/3$ before the radiation-dominated stage. 

It is already known that the simplest instantaneous transition leads to UV divergence which is at least logarithmic, and further regularizations should be imposed to get a finite total power of PGWs. 
However, an instantaneous transition is unphysical.  Recent lattice simulations show that typical duration of the transition is $\Delta\eta\sim1/\calH$ \cite{Lozanov:2016hid,Lozanov:2017hjm,Antusch:2020iyq,Saha:2020bis}. 
In this Section, we will show that a finite duration $\Delta\eta$ can suppress the UV part of the PGW spectrum exponentially, so no further regularization is needed.

\subsection{Inflation-radiation model}
\label{sub:inflation_radiation_model}

\begin{figure}[ht]
    \centering
    \begin{subfigure}[b]{.45\textwidth}
        \centering
        \includegraphics[width=\textwidth]{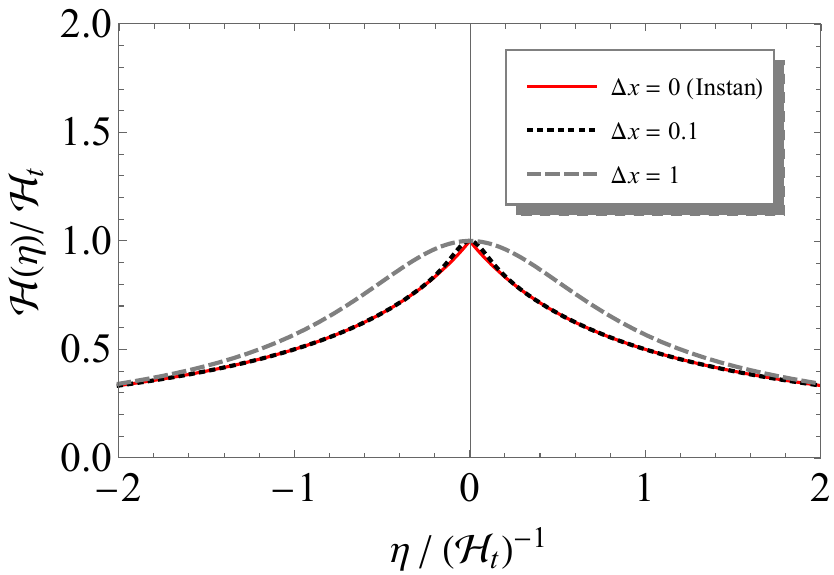}
    \end{subfigure}
    \hfill
    \begin{subfigure}[b]{.45\textwidth}
        \centering
        \includegraphics[width=\textwidth]{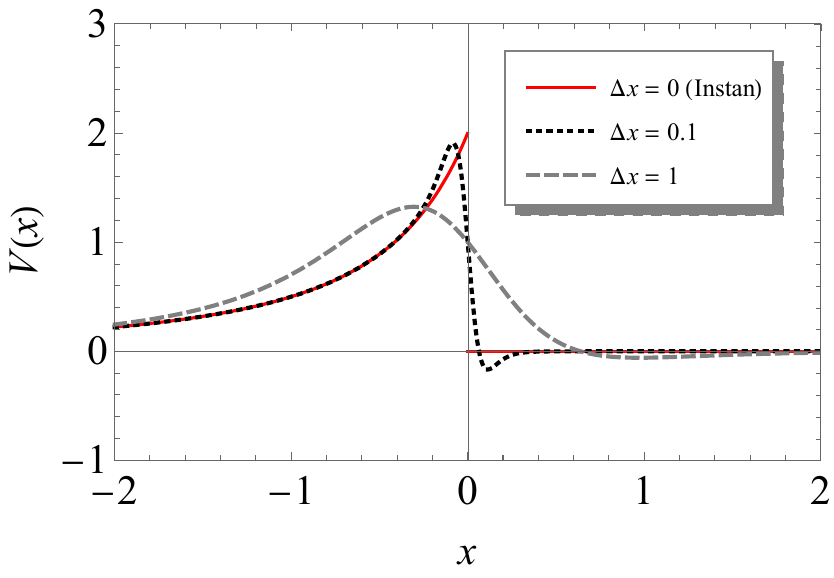}
    \end{subfigure}
    \hfill
    \caption{These figures show the evolutionary trends of the conformal Hubble parameter $\mathcal{H}(\eta)$ and the effective potential $V(x)$ in different $\Delta x$ cases, where the post-inflation stage is radiation-dominated.
    }
    \label{H&V}
\end{figure}

We choose the conformal Hubble parameter to describe the transition from inflation to the post-inflationary era, which means $\calH\to-1/\eta$ for $\eta\to-\infty$, and $\calH\to 1/\eta$ for $\eta\to+\infty$.\footnote{ 
We mention that the equation of motion for the canonical variable $t$ turns out to be the same in these two limits, dubbed the Wands duality \cite{Wands:1998yp}, as the only difference is the sign of $\calH$.} 
Here, we note that we have adopted the pure de Sitter approximation for the inflationary stage, which is known to be a very good approximation when dealing with the linear tensor perturbation. For an instantaneous transition, we can parameterize the conformal Hubble parameter as
\begin{align}\label{HC0}
	\mathcal{H}\equiv\frac{a'}{a}&=
	\begin{cases}
		-1/(\eta-\calH_t^{-1}) & \text{for} ~\eta<0,\\
  \\
		1/(\eta+\calH_t^{-1}) & \text{for} ~ \eta\geq0,
	\end{cases}
\end{align}
where $\calH_t\equiv a_tH_{\inf}$ is the conformal Hubble parameter at the transition $\eta=0$, with $a_t$ the scale factor at the transition. The derivative of $\calH$ is discontinuous, which causes the UV divergence in the tensor perturbation. To take into account the finite duration of the transition, we use the following parameterization of $\calH$ that gives a smooth $\calH$ and appropriately realizes 
the two limits of \eqref{HC0} for $\eta\to\mp\infty$,
\begin{equation}\label{conformal Hubble}
\cH\equiv\frac{a'}{a}=\frac1{\displaystyle \eta\tanh\left(\frac\eta{\Delta\eta}\right)+\cH_t}.
\end{equation}
The new parameter $\Delta\eta$ now describes the time scale of the transition process, as displayed in Fig.~\ref{H&V}. It is determined by the shape of the inflation potential, the inflaton decay rate, the time duration of thermalization of the created particles, etc. As displayed in Fig.~\ref{H&V}, \eqref{conformal Hubble} reduces to \eqref{HC0} in the limit $\Delta\eta\to0$. 
Here we note that, provided that we fix the scale factor at sufficiently later times to have the same normalization (i.e., $a(\eta)$ is the same at $\eta\gg \calH^{-1}$ independent of $\Delta\eta$), the peak value of the conformal Hubble parameter $\calH_t$ depends on $\Delta\eta$.

For convenience, we introduce a dimensionless time variable $x\equiv\calH_t\eta$ and a dimensionless momentum $\kappa\equiv k/\calH_t$.  Following \eqref{eomvp}, the mode function $v_\kappa(x)=t_k(\eta)/a(\eta)$ satisfies 
\begin{equation}\label{evolution equation}
	\frac{d^2v_\kappa}{dx^2}+(\kappa^2-V(x))v_\kappa=0,
\end{equation}
where 
\be\label{relation between V and H}
V(x)=\frac{a''}{a\calH_t^2}=\frac{\calH'+\calH^2}{\calH_t^2}=\frac{\displaystyle \tanh{\frac x{{\Delta x}}+\frac{x}{{\Delta x}}\left(1-\tanh^2\frac x{{\Delta x}}\right)-1}}{\displaystyle \left(x\tanh\frac x{{\Delta x}}+1\right)^{2}},
\ee
with $\Delta x=\calH_t\Delta\eta$. 
This effective potential is typical, and the following discussion can be easily generalized to any alternative parameterization of \eqref{HC0} which makes $\calH$ continuous up to its derivative. It is obvious that the asymptotic behaviors of $V(x)$ coincide with the instantaneous case,
\begin{align}\label{VC0}
	V(x)&=
	\begin{cases}
		2/(x-1)^2 & \text{for} ~ x<0 ~\text{and} ~|x|\gg \Delta x,\\
  \\
		0 & \text{for} ~ x>0~\text{and}~ |x|\gg \Delta x.
	\end{cases}
\end{align}

Eq.\eqref{evolution equation} is reminiscent of the stationary Schr\"{o}dinger equation in the one-dimentional barrier penetration problem, with $V(x)$ being the potential of the barrier. Except for some special cases, for instance, the P\"{o}schl-Teller potential, such an equation cannot be solved analytically. However, for our purpose, it is sufficient to obtain an approximate analytical solution in the UV frequency band, i.e. $\kappa^2\gg V(x)$ (roughly $k\gg \text{max}(\mathcal{H}_t)$), which allows us to treat $V(x)$ as a perturbation to the homogeneous equation of $d^2v/dx^2+\kappa^2v=0$. To the leading order, the two independent solutions are
 \begin{equation}\label{homogeneous solution}
     v_{0\kappa}(x)=\frac{1}{\sqrt{2\kappa}}e^{-i\kappa x}, \qquad \bar{v}_{0\kappa}(x)=\frac{1}{\sqrt{2\kappa}}e^{i\kappa x}.
 \end{equation}
The higher-order solutions can be obtained by the Green function method. The exact solution to \eqref{evolution equation} can be expressed in the integral form,
 \begin{equation}\label{exact solution}
     v_\kappa(x)=v_{0\kappa}(x)+\int_{-\infty}^{\infty}G_{\kappa}(x;\tilde{x})V(\tilde{x})v_{0\kappa}(\tilde{x})d\tilde{x},
 \end{equation}
where $G_k$ is the Green function for \eqref{evolution equation} with the retarded boundary condition. With the perturbation expansion of the Green function, the solution can be iteratively obtained order by order. The first-order approximation of the Green function is 
\begin{align}\label{Born Green}
    G^{(0)}_{\kappa}(x;\tilde{x})&\equiv\frac{\bar{v}_0(x)v_0(\tilde{x})-v_0(x)\bar{v}_0(\tilde{x})}{\bar{v}_0'(\tilde{x})v_0(\tilde{x})-v_0'(\tilde{x})\bar{v}_0(\tilde{x})}\Theta(x-\tilde{x})=\frac{1}{\kappa}\sin{\kappa(x-\tilde{x})}\Theta(\tilde{x}-x),
\end{align}
where $\Theta(x)$ is the Heaviside step function. To estimate the error, we notice that $G^{(0)}_{\kappa}(x;\tilde{x})$ is related to $G_{\kappa}(x;\tilde{x})$ by
\begin{align}
    G_{\kappa}(x;\tilde{x})&=G^{(0)}_{\kappa}(x;\tilde{x})+\int_{-\infty}^{\infty} d y G_{\kappa}(x;y)V(y)G^{(0)}_{\kappa}(y;\tilde{x})\notag\\
    &=G^{(0)}_{\kappa}(x;\tilde{x})+G_{\kappa}(x;y')\int_{-\infty}^{\infty} d y V(y)G^{(0)}_{\kappa}(y;\tilde{x}),
\end{align}
where $y'\in(x,x')$ is determined by the Lagrangian mean value theorem. Then the difference between the Green function and its leading-order approximation can be estimated as
\begin{equation}
    \int_{-\infty}^{\infty} d y V(y)G^{(0)}_{\kappa}(y;\tilde{x})\approx \mathcal{O}\left(\frac{V\Delta x}{\kappa}\right),\notag
\end{equation}
which is negligibly small in the UV regime we are interested in.

The solution $v_\kappa(x)$ to \eqref{evolution equation} has an asymptotic form when $x\gg1$ (i.e. $\eta\gg\Delta\eta$),
\begin{equation}\label{stable solution}
v_\kappa(x) = \alpha_\kappa v_{0\kappa}(x) + \beta_\kappa \bar{v}_{0\kappa}(x),
\end{equation}
where $\alpha_\kappa$ and $\beta_\kappa$ are the Bogoliubov coefficients discussed in \ref{sub:initial_condition_and_anlytical_solution}.
By substituting \eqref{Born Green} into \eqref{exact solution}, we obtain
\begin{equation}\label{coefficient}
    \begin{split}
        \alpha_\kappa &= 1+\frac{i}{2\kappa}\int_{-\infty}^{\infty} V(x')dx'\left[1+\mathcal{O}\left(\frac{\max(V(x))\Delta x}{\kappa}\right)\right],\\
        \beta_\kappa &= -\frac{i}{2\kappa}\int_{-\infty}^{\infty} e^{-2i\kappa x'}V(x')dx'\left[1+\mathcal{O}\left(\frac{\max(V(x))\Delta x}{\kappa}\right)\right].
    \end{split}
\end{equation}
To compute the PGW spectrum, we first evaluate the following:
\begin{align}
&|\alpha_\kappa|^2 = 1 - 2\ImPart\left[\frac{1}{2\kappa}\int_{-\infty}^{\infty} V(x')dx' \mathcal{O}\left(\frac{\max(V(x))\Delta x}{\kappa}\right)\right]+ \left(\frac{1}{2\kappa}\int_{-\infty}^{\infty} V(x')dx'\right)^2,\\
&|\beta_\kappa|^2 = \left(\frac{1}{2\kappa}\int_{-\infty}^{\infty} e^{-2i\kappa x'}V(x')dx'\right)^2\left(1+\mathcal{O}\left(\frac{\max(V(x))\Delta x}{\kappa}\right)\right).
\end{align}
The above integrals satisfy
\begin{equation}
\int_{-\infty}^{\infty} V(x')dx'\gtrsim \int_{-\infty}^{\infty} e^{-2i\kappa x'}V(x')dx'\,,
\end{equation}
which means that $|\beta_\kappa|^2$ is of the same order or smaller than the second term of $|\alpha_\kappa|^2$. 
Thus one might think it would be necessary to calculate $\alpha_\kappa$ up to next-to-leading order. 
However, as we have already seen in \eqref{Ogw}, since the GW energy spectrum is proportional to $|\beta_k|^2$, the leading-order $\beta_\kappa$ is enough. In fact, because of the identify $|\alpha_\kappa|^2=|\beta_\kappa|^2+1$ that holds for any Bogoliubov coefficients, computing $|\beta_\kappa|^2$ is enough to obtain $|\alpha_\kappa|^2$ to the same accuracy.

The integral for $\beta_\kappa$ in \eqref{coefficient} can be evaluated by taking a contour integral which goes clockwise around an infinitely large semi-circle on the lower half complex $x$-plane and then goes along the real axis, as the integrand $V(x)e^{-2i \kappa x}$ decays exponentially when  $|x|\to\infty$ on the lower half complex plane. 
Therefore the integral in \eqref{coefficient} is given by the sum of residues of all the poles on the lower half complex plane. In the UV limit ($\kappa\gg1$), the dominant pole $x_p$ is the one closest to the real axis (i.e., the one with the smallest imaginary part in absolute magnitude), which is proportional to the decay rate of the PGW spectrum. Considering the coefficients of the short-wavelength modes, we have
\begin{equation}\label{beta factor}
    \beta_\kappa \approx \frac{\pi}{\kappa} \lim_{x\to x_p}\left[\frac{(e^{-2i\kappa x}V(x)(x-x_{p})^{n_{p}})^{[n_{p}-1]}}{(n_{p}-1)!}\right]\propto \kappa^{(n_p-1)} e^{-\mu \kappa/2},
    \qquad\mathrm{with}~\mu\equiv-4\left|\ImPart\left(x_p\right)\right|. 
\end{equation}
where the superscript $[n_p-1]$ means the $(n_p-1)$-th derivative and $n_p$ is the order of the pole $x_p$. The detailed derivation of \eqref{beta factor} is given in Appendix \ref{general_potential}. 
For our analytical model \eqref{relation between V and H}, we find the dominant pole of $V(x)$ is the same as that of $\cH$. 
We have no rigorous proof, but it seems reasonable to assume that such a pole of $\calH(\eta)$ in the complex plane exists if $\calH(\eta)$ is a smooth (i.e., analytic) function that connects the asymptotic behaviors of inflation and post-inflationary stages, say \eqref{HC0}. Substituting \eqref{beta factor} into \eqref{Ogw}, we find that $\mu$ is the decay rate of the energy spectrum, which is determined by $\eta_t$ and $\Delta\eta$. 

\begin{figure}
    \centering
    \begin{subfigure}[b]{.44\textwidth}
	\centering
        \includegraphics[width=\textwidth]{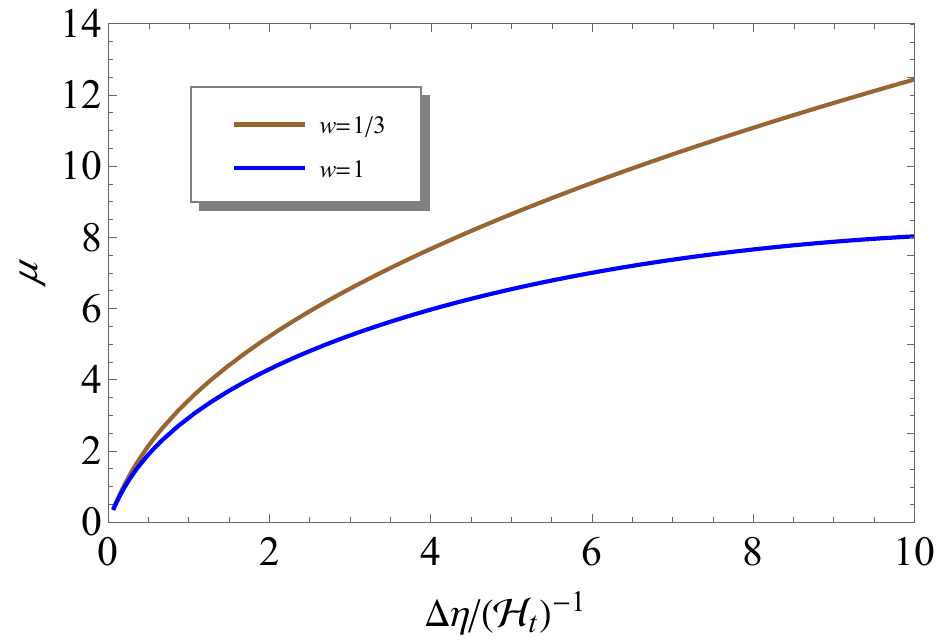}
    \end{subfigure}
    \hfill
    \begin{subfigure}[b]{.46\textwidth}
	\centering
	\includegraphics[width=\textwidth]{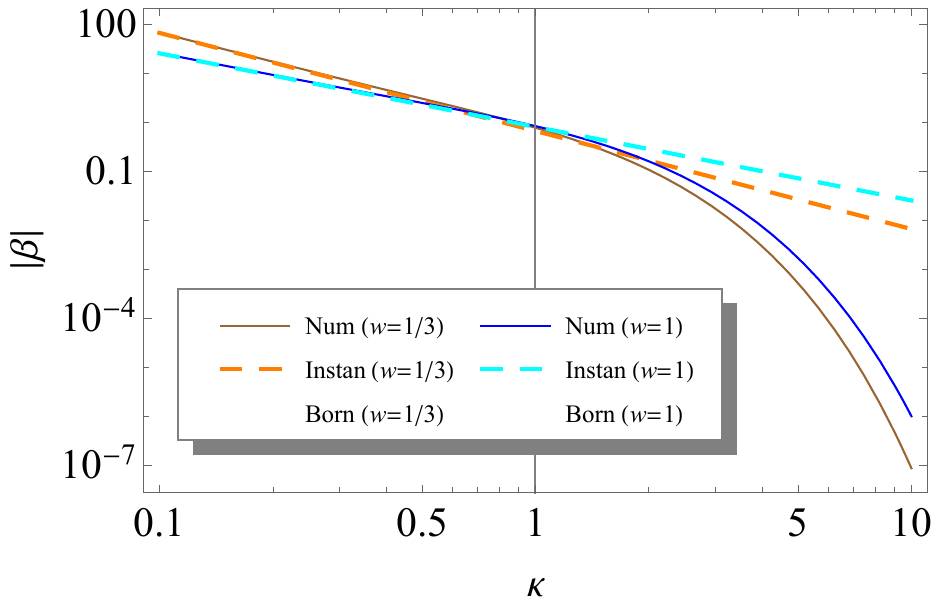}
    \end{subfigure}
    \caption{\textbf{[Left]} The decay rate $\mu$ as a function of transition time $\Delta\eta$, for $w=1/3$ (red, solid) and $w=1$ (blue, dashed). \textbf{[Right]} The coefficient $|\beta_\kappa|$ given by the Born approximation (red), the instantaneous approximation (cyan) and the numerical calculation (blue) as a function of the wavenumber $\kappa=k/\calH_t$, where $\calH_t$ is the conformal Hubble parameter at the time of transition. The solid lines are  for the inflation-radiation model and the dotted lines are for the inflation-stiff-EoS ($w=1$) model. The transition time scale is set to ${\Delta x}=\calH_t\Delta\eta$=1. 
    }
    \label{fig:mu-dt & beta}
\end{figure}

For our specific effective potential $V(x)$ on \eqref{relation between V and H}, the poles are solutions of the transcendental equation $\calH^{-1}(x)=0$ where $\calH$ is given by \eqref{conformal Hubble}. 
It is easy to see that the dominant pole $x_p$ is on the imaginary axis in the range $-\frac{\pi}{2}\Delta x<\ImPart(x_p)<0$, which is shown explicitly in Appendix \ref{coefficients}. The Bogoliubov coefficient given by \eqref{coefficient} is now
\begin{align}
	\label{rad beta factor}
	\beta_\kappa &=-\frac{2\pi(R_\text{r}(i x_p/{\Delta x})-i) e^{-\mu\kappa/2}}{R_\text{r}(i x_p/{\Delta x})^2}+\mathcal{O}(\kappa^{-2})\,;
 \quad R_\text{r}(z)=\tan{z}+\frac{z}{\cos^2{z}},
\end{align}
which is consistent with the numerical result when $\kappa\gg1$, shown by the right panel of Fig.~\ref{fig:mu-dt & beta}. It is obvious that the decay rate only depends on the product  $\calH_t\Delta\eta$ (i.e. ${\Delta x}$) in our parameterization, which is shown by the left panel of Fig.~\ref{fig:mu-dt & beta}. 
For comparison, we also plot $|\beta_\kappa|$ under the instantaneous transition approximation in the right panel of Fig 2. It is shown that the IR behavior is independent of $\Delta\eta$, which is the same for the instantaneous transition. This is shown explicitly in Appendix~\ref{IRtail}. 

\begin{figure}
    \centering
    \begin{subfigure}[b]{.45\textwidth}
		\centering
		\includegraphics[width=\textwidth]{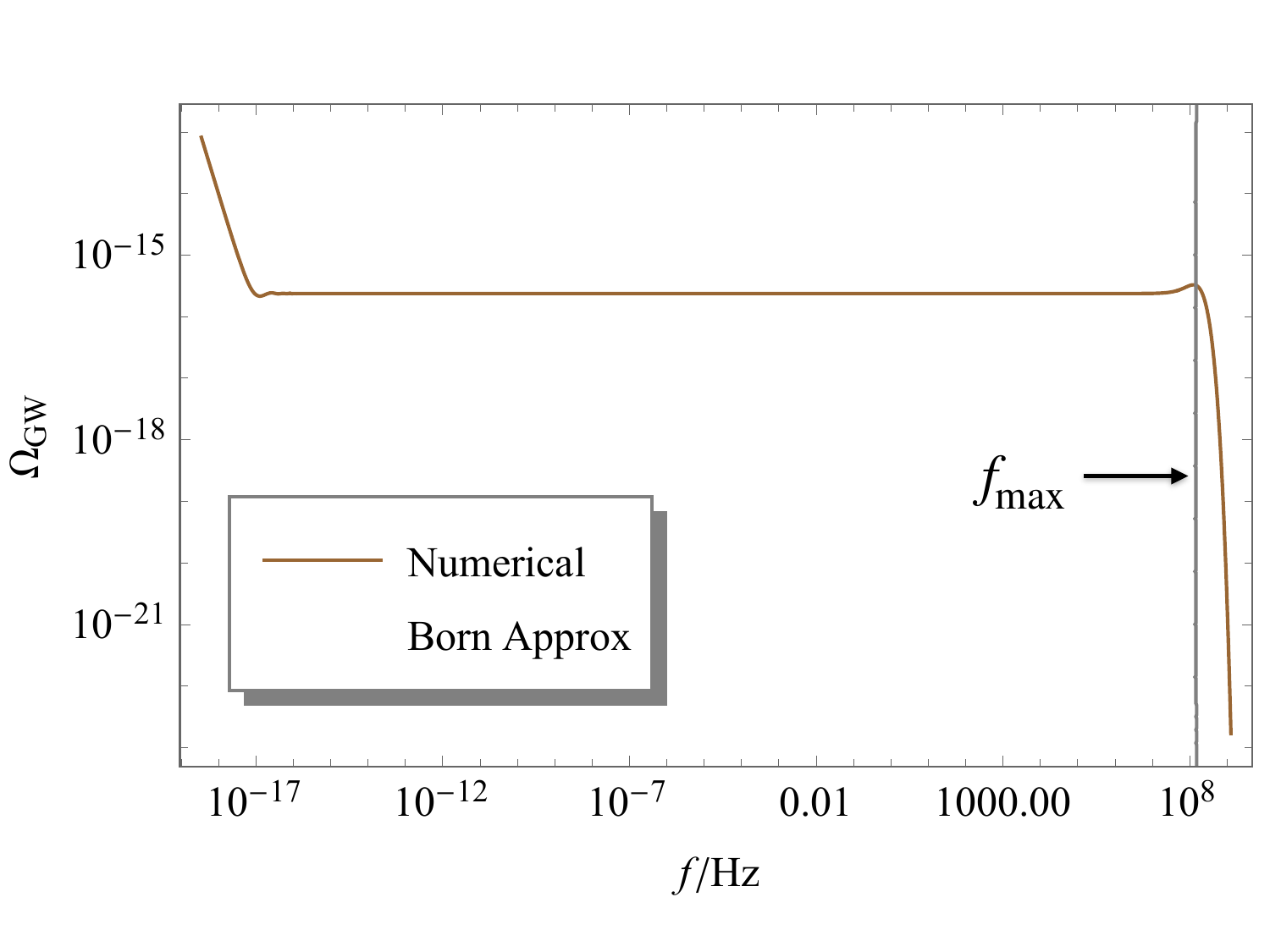}
	\end{subfigure}
	\hfill
	\begin{subfigure}[b]{.45\textwidth}
		\centering
		\includegraphics[width=\textwidth]{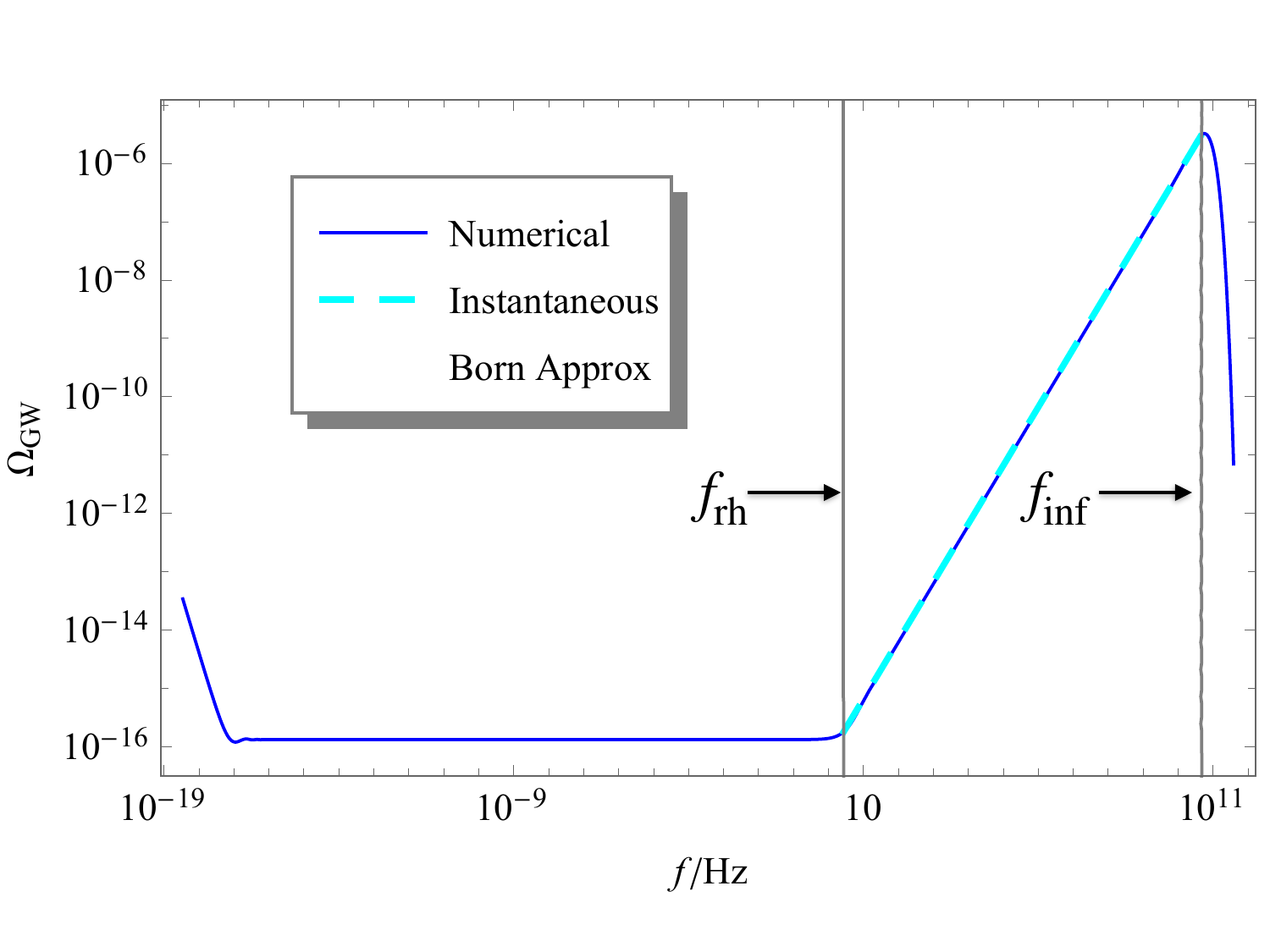}
	\end{subfigure}
	\hfill
    \caption{The PGW energy spectrum in the inflation-radiation model (left), and in the inflation-$w$ model with $w=1$ and $T_\rh=10^8 ~\text{GeV}$ (right), together with $r=0.036$ and $\Delta x=1$.}
    \label{fig: PGW spectrum}
\end{figure}

In order to convert the dimensionless momentum into the physical frequency observed today, we note that $f_0={k c}/({2\pi a_0})$ and $k=\kappa\calH_t$. The critical frequency above which the modes always stay subhorizon is given by $f_{\rh}={\calH_t c}/({2\pi a_0})$. 
After reaching the peak, the GW energy spectrum decays as 
\begin{equation}\label{Rad PS}  
	\Omega_{\GW}(f>f_\rh)\propto f^4 e^{-\mu f/f_{\rh}}\,;\quad \mu=4|\ImPart(x_p)|\,,
\end{equation}
shown in the left panel of Fig.~\ref{fig: PGW spectrum}.
The decay rate of the GW spectrum $\mu$ is also related to the transition time ${\Delta x}$ via $x_p$. This means that the UV behavior of the GW spectrum, if observed, can give a strong constraint on reheating models.
 Furthermore, this exponential decay is model-independent as discussed in Appendix \ref{general_potential}, while the power-law index of the prefactor depends on the order of the dominant pole, which depends on models. 
 If the dominant pole is of $n_p$-th order, the prefactor in \eqref{Rad PS} becomes $k^{4+2(n_p-1)}$.
 Eq.\eqref{Rad PS} clearly shows that instantaneous reheating gives $\mu\propto\Delta x\to0$, hence generates a UV divergent GW spectrum.
  
The physical frequency $f$ we observe today is related to wavenumber $k$ by \cite{Watanabe:2006qe}
\begin{align}\label{end of reheating}
	f=\frac{1.66 \times 10^{-7}}{2\pi}\left(\frac{T_k}{1\textrm{GeV}}\right)\left[\frac{g_{*s}(T_k)}{106.75}\right]^{-1/3}\left[\frac{g_{*}(T_k)}{106.75}\right]^{1/2}\textrm{Hz},
\end{align}
where $T_k$ is the temperature at the horizon crossing. The Hubble parameter during inflation is given by the power spectrum of the tensor perturbation,
\begin{equation}\label{eq:tensor spectrum}
    \mathcal{P}_T = \frac{2H^2_{\inf}}{\pi^2\mpl^2},
\end{equation}
where $\mathcal{P}_T$ is related to the spectrum of the curvature perturbation by the tensor-to-scalar ratio, $\mathcal{P}_T=r\mathcal{P}_{\calR}$.
For a typical (high-scale) model of inflation, the temperature of the radiation-dominated universe right after the end of inflation is \cite{Liddle:2003as,Dai:2014jja,Creminelli:2014oaa,Gong:2015qha,Cai:2015soa,Mishra:2021wkm}
\begin{align}
T_\mathrm{max}\approx\frac{90\mpl^2 H_{\inf}^2}{\pi^2 g_*}= 5.78 \times10^{15}\left[\frac{g_*}{106.75}\right]^{-1/4}\left[\frac{r}{0.036}\right]^{1/4}\left[\frac{\mathcal{P}_\calR}{2.09\times10^{-9}}\right]^{1/4}\mathrm{GeV},
\end{align}
where we have neglected the change of the Hubble parameter during inflation. By \eqref{end of reheating}, we can get the highest frequency of the plateau of the GW spectrum in the high scale models,
\be\label{fmax}
f_\mathrm{max}=1.53\times 10^{8} \left[\frac{g_{*s}(T_k)}{106.75}\right]^{-1/3}\left[\frac{g_{*}(T_k)}{106.75}\right]^{1/4}\left[\frac{r}{0.036}\right]^{1/4}\left[\frac{\mathcal{P}_\calR}{2.09\times10^{-9}}\right]^{1/4}\mathrm{Hz},
\ee
which is shown in the left panel of Fig.~\ref{fig: PGW spectrum}.

\subsection{Stiff Equation of State} 
\label{sub:stiff_equation_of_state}
The discussion above can be generalized to an arbitrary eEoS parameter $w$. Especially, if there is a post-inflationary era with $w>1/3$ before the radiation-dominated era, the GW energy spectrum will be enhanced as the background energy density decays faster than radiation, which makes it easier to be detected. 
Suppose the inflation ends at $a_\mathrm{inf}$, and the $w$-dominated era ends at $a_\mathrm{rh}$, after which the universe becomes radiation-dominated. The EoS can be parameterized as 
\begin{equation}
	w=\left\{
		\begin{matrix}
			w_i && a<a_{\inf},\\
			w &&  a_{\inf}<a<a_{\rh},\\
			1/3 && a_{\rh}<a<a_{\mathrm{eq}},\\
			0 && a_{\mathrm{eq}}<a.
		\end{matrix}
	\right.
\end{equation}
For simplicity, we assume $w_i=-1$ during inflation. The $\Omega_\mathrm{GW}$ with wavenumbers reentering the horizon during the $w$-dominated era has a power index of $n_{\GW}=n_T+2(3w-1)/(3w+1)$~\cite{DEramo:2019tit}, where $n_T$ is the primordial tensor spectral tilt. Similar to the discussion in Sec.~\ref{sub:inflation_radiation_model}, for the transition at the end of inflation, we can generalize the smooth transition \eqref{conformal Hubble} to
\begin{align}\label{stiff Hubble}
\cH(\eta)=\frac1{\displaystyle \left(\frac{3w+3}{4}\tanh\frac\eta{\Delta\eta}+\frac{3w-1}{4}\right)\eta+\cH_t^{-1}}.
\end{align}

For the effective potential $V(x)$ generated by the above $\calH(\eta)$, the poles are solutions of the transcendental equation $\calH^{-1}(\eta)=0$. Similarly to the radiation case, the pole closest to the real axis, $x_p$, dominates the contour integral. Parallel to the calculations in the previous subsection, the Bogolyubov coefficient $\beta_\kappa$ follows the same formula \eqref{rad beta factor}, where the residue function $R_\text{r}$ should be generalized to
\begin{equation}
    R_w(z)=\frac{1}{z{\Delta x}}+\frac{3z(w+1)}{4\cos^2z}\,,
\end{equation}
and the decay rate formula remains the same, $\mu=4|\text{Im}(x_p)|$. 
It is shown in the right panel of Fig.~\ref{fig:mu-dt & beta} that our method fits well with the numerical result when $\kappa\gg1$ in the $w$-dominated post-inflation era. The IR behavior remains independent of $\Delta \eta$, indicating that the instantaneous transition approximation still holds. For details, see Appendix~\ref{IRtail}.

\begin{figure}[ht]
    \centering
    \begin{subfigure}[b]{.45\textwidth}
        \centering
        \includegraphics[width=\textwidth]{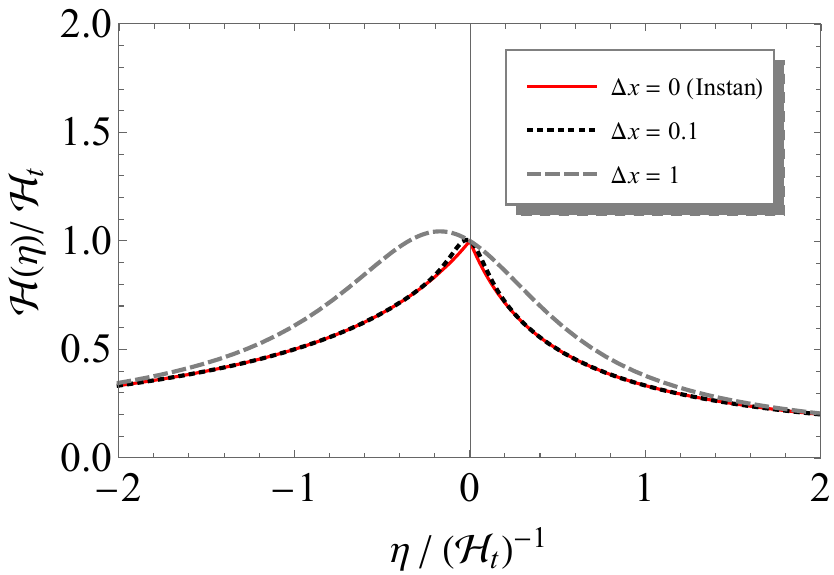}
    \end{subfigure}
    \hfill
    \begin{subfigure}[b]{.45\textwidth}
        \centering
        \includegraphics[width=\textwidth]{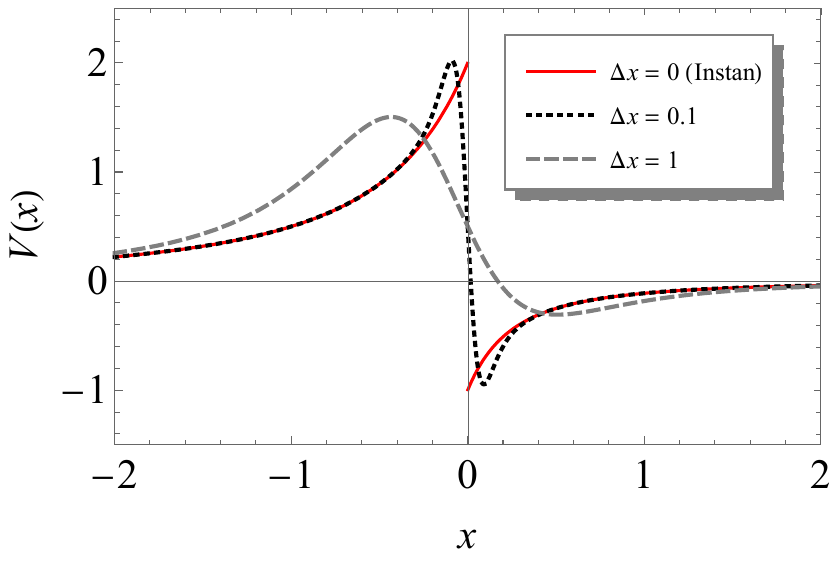}
    \end{subfigure}
    \hfill
    \caption{These figures, which have similar profiles as Fig.\ref{H&V}, show the evolutionary trends of the conformal Hubble parameter $\mathcal{H}(\eta)$ and the effective potential $V(x)$ in the inflation-$w$-era model
    }
    \label{Stiff H&V}
\end{figure}

In this $w$-dominated case, $\Omega_\mathrm{GW}$ has a few characteristic features, shown in the right panel of Fig.~\ref{fig: PGW spectrum}. It starts to increase from $f_\rh$ and reaches the maximum at $f_\mathrm{inf}$.
The former frequency corresponds to the mode which re-enters the horizon at the onset of the radiation-dominated era, while the latter has the same wavelength as the Hubble horizon at the end of inflation. 
The relation between $f_{\inf}$ and $f_\rh$ is
\begin{align}\label{kend}
	f_{\inf}&=\frac{a_{\inf}H_{\inf}}{2\pi a_0}=\frac{a_{\inf}H_{\inf}}{a_{\rh}H_{\rh}}f_{\rh},
\end{align}
where $H_{\rh}$ is the Hubble parameter when the universe becomes radiation-dominated. By using \eqref{end of reheating}, \eqref{kend} becomes
\begin{align}\label{def:finf}
    f_{\inf} &=1.53\times10^{8} \text{Hz}~ \left[\frac{g_{*s}(T_\rh)}{106.75}\right]^{-1/3}\left[\frac{g_{*}(T_\rh)}{106.75}\right]^{\frac{1}{3(1+w)}}\notag\\
    &\times\left[\frac{r}{0.036}\times\frac{\mathcal{P}_\calR}{2.09\times10^{-9}}\right]^{\frac{1}{2}-\frac{1}{3(1+w)}}\left[\frac{T_\rh}{5.78\times10^{15}\text{GeV}}\right]^{\frac{4}{3(1+w)}-1} ,
\end{align}
which reduces to \eqref{fmax} when $w=1/3$.

The energy spectrum between $f_\mathrm{rh}$ and $f_\mathrm{inf}$ is
\begin{equation}\label{Power law}
    \Omega_{\GW}(f_{\inf}>f>f_{\rh})=\Omega_{\GW}^{(\mathrm{flat})}\left(\frac{f}{f_{\rh}}\right)^\frac{2(3w-1)}{(3w+1)},
\end{equation}
where $\Omega_{\GW}^{(\mathrm{flat})}$ is the spectral amplitude of those modes which reenter the horizon during the radiation dominated era, determined by the tensor-to-scalar ratio $r$ and the primordial tensor tilt $n_T$. The UV spectrum with $f>f_{\inf}$ displays an exponential decay,
\begin{equation}
	\Omega_\GW(f>f_{\inf})\propto f^4 e^{-\mu f/f_{\inf}}\,;\quad \mu=4|\ImPart(x_p)|.
\end{equation}
As we mentioned before, the definition of $x_p$ is generalized to the $w$-dominated case without any changes in the expression for $\mu$.

\section{Fitting formula and Its application}
\label{Fitting formula}

Except for the UV and IR limits, the PGW spectrum cannot be expressed in terms of a simple function. However, our numerical result shows that it can be well approximated by a fitting formula,
\begin{equation}\label{fitform}
	\Omega_{\GW}(f)=A \left(\left(\frac{f}{f_{p}}\right)^{n_\mathrm{GW}}\frac{1}{\cosh(f/f_d)}+p\left(\frac{f}{f_{p}}\right)^{4}e^{-\mu_* (f/f_{p}-1)}\right),
\end{equation}
where $n_\mathrm{GW}=n_T+2(3w-1)/(3w+1)$, $\mu_*$ denotes the fitted decay rate, and $f_p$ denotes the fitted frequency of the peak. The hyperbolic cosine function is used to remove the power law contribution on the UV side.  
We mention that the above fitting formula is valid for not-too-slow transitions, because the second term in Eq.~\eqref{fitform} would dominate the IR part, $f<f_p$, if $\mu_*\propto\calH_t\Delta\eta\gg1$. 
Therefore we only focus on the case $\calH_t\Delta\eta \lesssim1$, which seems reasonably general. 

Here we note that there is a relation between the peak amplitude and the peak frequency. Roughly speaking, the amplitude of the tensor perturbation is $H_{\inf}/(\pi\mpl)$ during inflation. 
The peak frequency is determined by the mode that just touches the horizon at the end of inflation, $k_{\inf}= a_{\inf} H_{\inf}$, where $a_{\inf}$ is the scale factor at the end of inflation. 
After that, it decays as $1/a$ and the current energy spectrum around the peak can be estimated as
\begin{equation}
    \Omega_{\GW}(f_{\inf})\sim 2\cdot \frac{8\pi G}{3H^2_0}\frac{ k_{\inf}^2 \langle h_{\inf}^2(\eta_{0})\rangle}{32\pi G a_0^2}= \frac{ a_{\inf}^2 k_{\inf}^2 \langle h_{\inf}^2(\eta_{\inf})\rangle}{6H^2_0 a_0^4}= \frac{(2\pi f_{\inf})^4}{6\pi^2H_0^2 \mpl^2},
\end{equation} 
where $\langle h_{\inf}^2(\eta_{\inf})\rangle$ is the mean square amplitude of the mode $k=k_{\inf}$, given by $H_{\inf}^2/(\pi\mpl)^2$. 
A more precise estimate can be done by adopting the instantaneous transition for the IR modes. One finds
\begin{equation}\label{eq:A and f_p}
    A^{\rm inst}= \frac{ \left(2\pi f^{\text{inst}}_p\right)^4 ~\Gamma{(n+1)}^2 }{12\pi^3\mpl^2 H_0^2} \left(\frac{2}{x_s}\right)^{2n+1}=1.10 \times 10^{-16} \left(\frac{f^{\text{inst}}_p}{1.53\times10^{8} \mathrm{Hz}}\right)^4\left( \frac{ \Gamma(n+1)^2 (1+3w)^{2n+1}}{\pi}\right),
\end{equation}
where the superscript ``inst'' stands for the instantaneous transition. We assume $n_T=0$ for simplicity, and $n=3(1-w)/2(1+3w)$. For details, see Appendix~\ref{IRtail}, where we also find that the amplitude of the peak includes a $\mathcal{O}(1)$ factor related to the transition time. 

\begin{figure}
    \centering
    \includegraphics[width=0.8\linewidth]{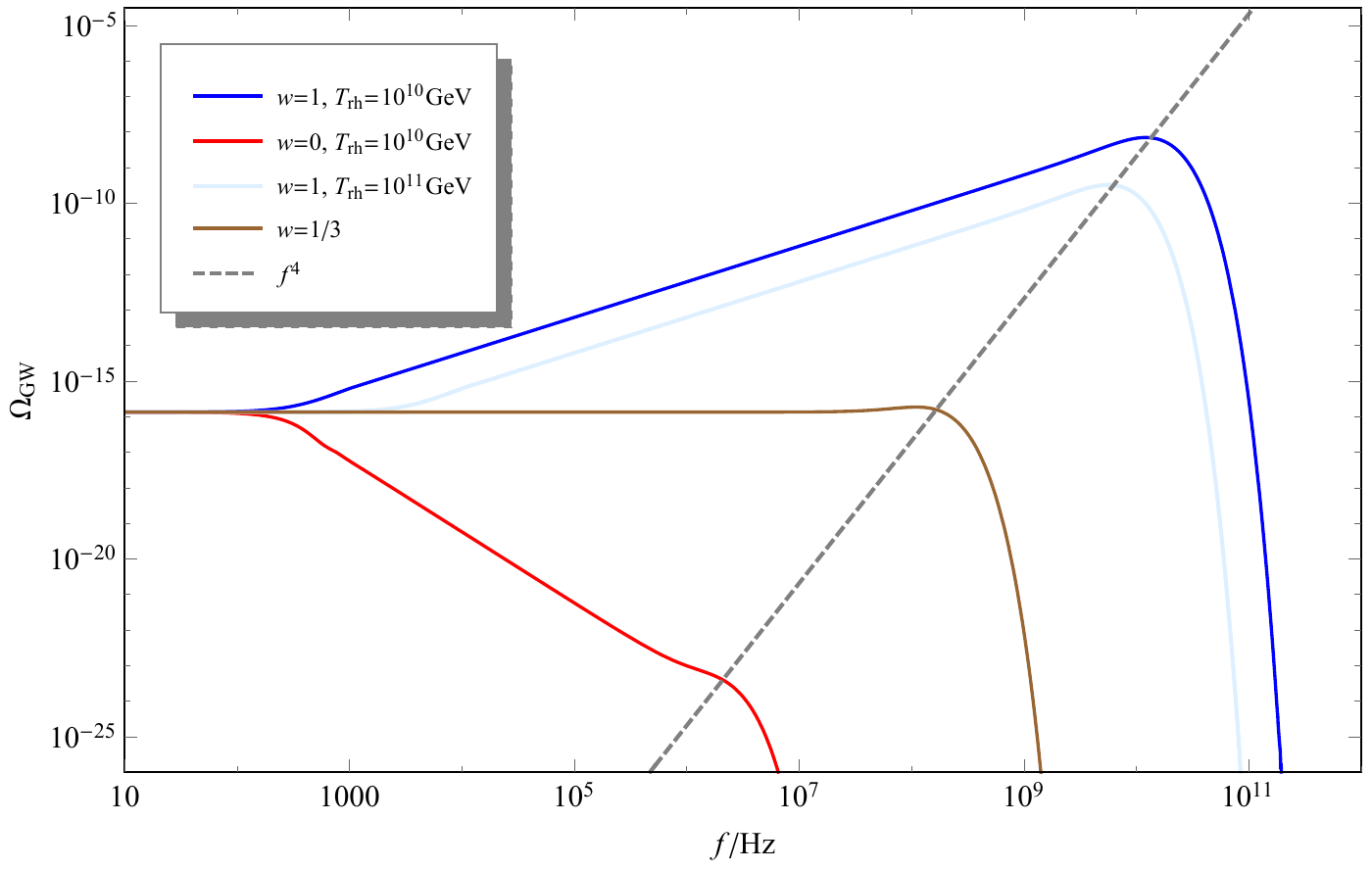}
    \caption{This figure shows that the peak relation \eqref{eq:A and f_p} (black dashed line) is universal for different parameters, shown in the box.
    } 
    \label{fig:peakrelation}
\end{figure}

Equation~\eqref{eq:A and f_p} relates the frequency and the amplitude at the peak, which is a direct consequence of the equation of motion \eqref{eomvp}, hence valid for any $-1/3<w\leq 1$. 
As shown in Fig.~\ref{fig:peakrelation}, except for an insignificant $w$-dependent factor of $\mathcal{O}(1)$, the peak amplitude always lie on a straight line of $A\sim1.10 \times 10^{-16}\left(f_{\inf}/1.53\times10^{8} \mathrm{Hz}\right)^4$, regardless of the other parameters.  Thus, the peak amplitude-frequency relation \eqref{eq:A and f_p} can be used to distinguish PGWs from a stochastic GW background of other origins. 

\begin{figure}[htbp]
	\centering
	\begin{subfigure}[b]{.45\textwidth}
		\centering
		\includegraphics[width=\textwidth]{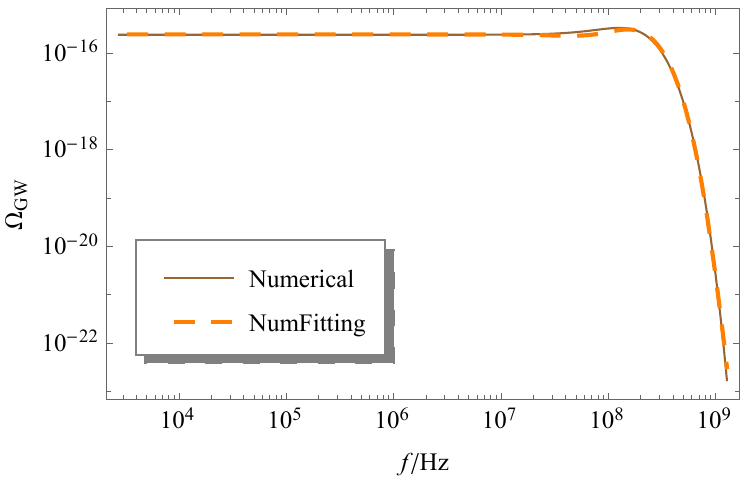}
	\end{subfigure}
	\hfill
	\begin{subfigure}[b]{.45\textwidth}
		\centering
		\includegraphics[width=\textwidth]{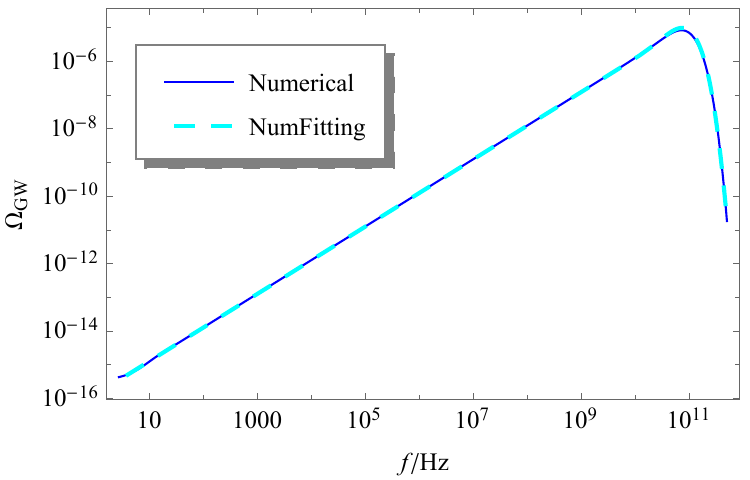}
	\end{subfigure}
	\hfill
	\caption{In these two figures we compare the fitting formula \eqref{fitform} with the numerical result in the inflation-radiation (left) and inflation-$w$-era model (right) with the same parameters used in Fig.\ref{fig: PGW spectrum}. For the left panel, the fitted parameters $A=10^{-15.6}$, $f_p=0.95 f_{\rh}$ and $\mu_*=1.02\mu$. For the right panel, they has $A=10^{-5.12}$, $f_p=1.00 f_{\inf}$ and $\mu_*=1.07 \mu$. The parameters we fitted also satisfactorily meet the peak relation \eqref{eq:A and f_p}.}
    \label{directfit}
\end{figure}

Now returning to the fitting formula \eqref{fitform}, we compare it with the GW energy spectrum calculated by numerically solving the equation of motion \eqref{evolution equation}, as is shown in Fig.~\ref{directfit}.
To fit the numerical result with \eqref{fitform}, we regard $A$, $f_d$, $p$, $f_p$, and $\mu_*$ as fitting parameters, whose physical meanings are the amplitude, the cutoff frequency of the IR power-law tail, the ratio between the IR and UV components, the frequency of the peak, and the UV decay rate, respectively.
Since $w$ can be easily determined by the slope of the IR side, it is regarded as the model parameter rather than a fitting parameter. 
Apparently there are too many free parameters that may harm the simplicity of formula \eqref{fitform}. 
After preliminary fittings, we find the relations $f_d\approx f_{\inf}/2$ and $p\approx 1$ are always satisfactory within 10\% accuracy. 
Thus, we will just set $f_d=f_{p}/2$ and $p=1$ in \eqref{fitform}. The remaining parameters are the amplitude $A$, the peak frequency $f_p$, and the UV decay rate $\mu_*$. The curve with the fitted formula is displayed in Fig.~\ref{directfit}, which fits the numerical results quite well. 

\begin{figure}[htbp]
	\centering
	\begin{subfigure}[b]{.48\textwidth}
		\centering
		\includegraphics[width=\textwidth]{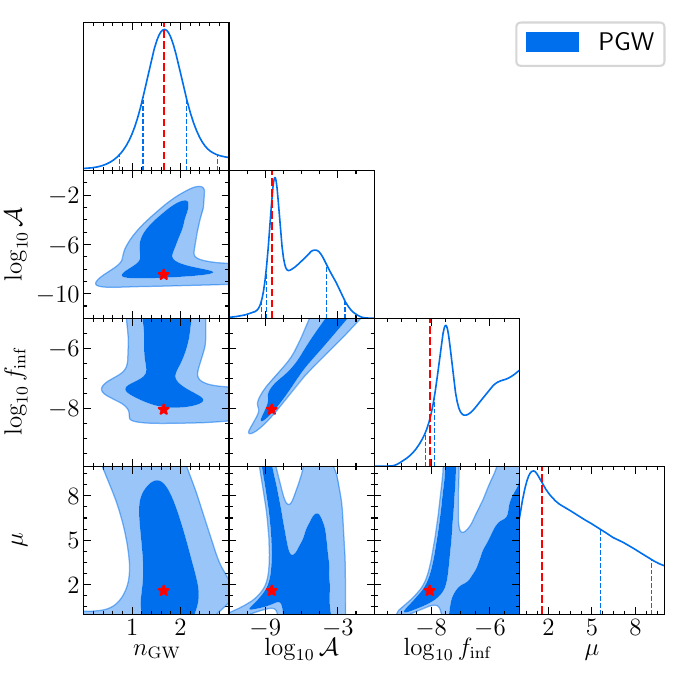}
	\end{subfigure}
	\hfill
	\begin{subfigure}[b]{.5\textwidth}
		\centering
		\includegraphics[width=\textwidth]{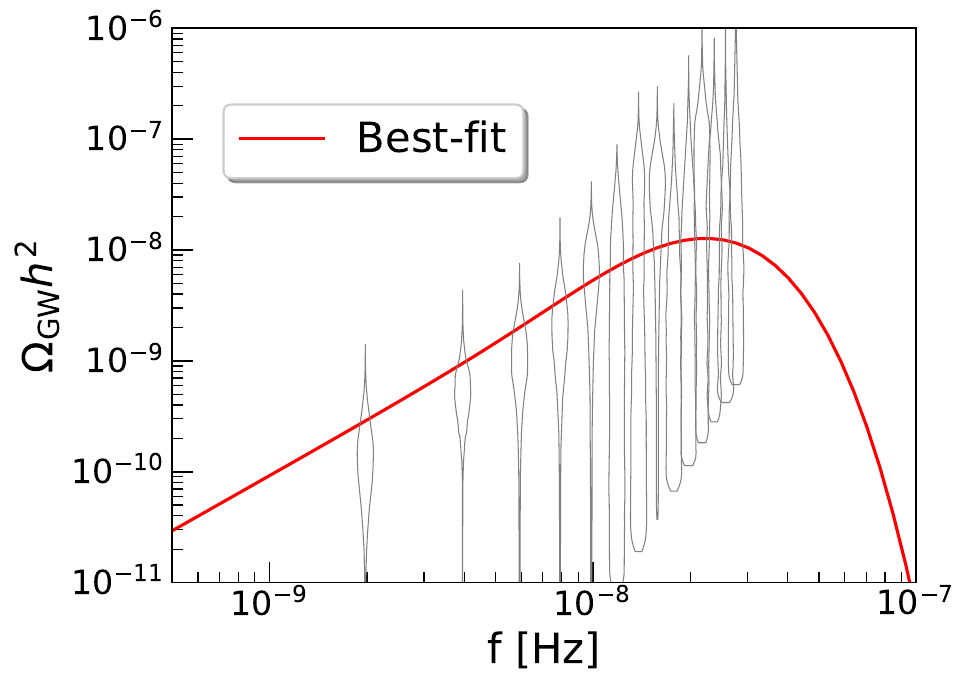}
	\end{subfigure}
	\hfill
	\caption{The Bayesian likelihood for the parameters in our fitting formula \eqref{fitform} is shown in the left panel together with the best-fit parameters: $n_\mathrm{GW}=1.65$, $\log_{10}A=-8.46$, $\log_{10}f_{\inf}=-8.05$, $\mu=1.59$ marked by red stars. The GW energy spectrum with the best-fit parameters is drawn on the right panel, together with the violin plots of the NANOGrav data.}
	\label{Fig: NANOGrav}
\end{figure}

As an example, let us apply our fitting formula to the NANOGrav dataset. Using the package \href{https://andrea-mitridate.github.io/PTArcade/}{\textsc{ptarcade}} in the ceffyl mode \cite{Lamb:2023jls}, we can intepret the nHz stochastic GW background as PGWs, of which the Bayesian likelihood contours of the model parameters are shown in Fig.~\ref{Fig: NANOGrav}. 
The two separate contours in the $\mu$-$\log_{10} f_p$ plot indicates that both terms in \eqref{fitform} can dominate the nHz GW spectrum. We also find the PGW interpretation requires a strongly blue-tiled spectrum $n_\mathrm{GW}\sim1.65$, aligning with Ref. \cite{NANOGrav:2023hde,Jiang:2023gfe,Ye:2023tpz}. 
However, in standard cosmology, we note that one would need a fine-tuning to make the peak feature appear within the PTA frequency band. Namely, the ratio of the conformal Hubble horizon at the end of inflation to that at the CMB epoch must be as low as $\mathcal{O}(10^5)$ and the GWs must be strongly enhanced compared to the adiabatic mode $h_{\inf}(\eta_{\inf})\sim 10^{16} H_{\inf}/(\pi \mpl)$.

\section{Conclusion and Discussion} 
\label{sec:conclusion}
In this paper, we revisited the UV divergence problem in PGWs. We clearly showed that it originates from the unphysical assumption of an instantaneous transition to a post-inflation era. 
By considering a finite duration of the transition, the PGW spectrum is found to decay exponentially towards high frequencies for the modes whose wavelengths never exceed the Hubble horizon, after the vacuum fluctuations in the UV limit are removed in a standard manner. 
Then we presented, for the first time, an analytical method to calculate the decay rate of the PGW spectrum, which agrees well with the numerical results. 
Assuming regularity of the Hubble parameter, the decay rate $\mu$ is determined by the imaginary part of the pole of $\calH$ closest to the real axis, where 
$\calH$ is the conformal Hubble parameter $\calH=aH$.

We also noted that the characteristic spectral shape of PGWs is a power law on the IR side and exponential decay on the UV side, and it is distinguishable from stochastic GW backgrounds from other sources. 
For instance, the spectrum of the scalar induced GWs from a lognormal peak also has an exponential decay on the UV side \cite{Pi:2020otn}, but it has a different IR scaling proportional to either $k^3$ or $k^2$, depending on the width of the scalar curvature perturbation spectrum \cite{Cai:2018dig,Cai:2019cdl}. 
This is different from the spectral index $n_\mathrm{GW}$ of PGWs with $n_T=0$, which is always smaller than 2, no matter what value of the EoS parameter $w$ is.\footnote{For induced GWs generated in a period with a different EoS parameter $w\neq1/3$, the IR scaling law may be smaller than 2.}
Besides the IR and UV behaviors, there exists a universal amplitude-frequency relation at the peak of the PGW spectrum \eqref{eq:A and f_p}, as long as the GW background originates from the adiabatic vacuum and is caused by changes in the vacuum state due to cosmic expansion. This relation can be used to identify the origin of the GW spectrum.

We also put forward a fitting formula \eqref{fitform}, which describes the PGW spectrum quite well around the peak of the spectrum, as shown in Fig.~\ref{directfit}. 
For a given set of parameters describing our cosmological model and the transition process, the fitting parameters we recover (particularly the decay rate $\mu$) are in good agreement with the values given by analytical calculations. This implies that Eq.~\eqref{fitform} is effective for data analysis when we detect such signals.
As an example, we used \eqref{fitform} to fit the recently detected nHz stochastic GW background. We found that, although the data can be fitted, as the signal does not satisfy the relation at the spectral peak, it cannot be of inflationary origin.\footnote{In non-inflationary or non-standard scenarios, there is a possibility that the spectrum is blue-tilted. We did not consider such possibilities in this paper.}
As is shown in Fig.~\ref{fig:peakrelation}, according to this universal amplitude-frequency relation, the peak in the PGW spectrum can only be probed by the high-frequency GW detectors \cite{Aggarwal:2020olq}.

\section*{Acknowledgement}
We thank Zhong-zhi Xianyu for useful comments and thank Bo Wang for helpful discussions. This work is supported by the National Key Research and Development Program of China Grant No.~2021YFC2203004, by Project 12047503 of the National Natural Science Foundation of China. It is also supported in part by JSPS KAKENHI Nos.~JP20K14461, JP20H05853 and JP24K00624, and by the World Premier International Research Center Initiative (WPI Initiative), MEXT, Japan.

\appendix
\section{General Effective Potential Discussion}\label{general_potential}
After substituting \eqref{Born Green} into \eqref{exact solution}, the first order approximation of the solution of \eqref{evolution equation} is equal to 
\begin{align}\label{Born} 
	v_\kappa(x)&=v_{0\kappa}(x)+\int_{-\infty}^{x}\frac{e^{i\kappa(x-x')}-e^{-i\kappa(x-x')}}{2\kappa i}V(x')v_{0\kappa}(x')\, dx'(1+\mathcal{O}(V(x)\Delta x/\kappa))\notag\\
	&=v_{0\kappa}(x)+\frac{i}{2\kappa}v_{0\kappa}(x)\int_{-\infty}^{x} V(x')\,dx'-\frac{i}{2\kappa}u_{0\kappa}(x)\int_{-\infty}^{x} e^{-2i\kappa x'}V(x')\,dx'+\mathcal{O}(\kappa^{-2})\,.
\end{align}
It is possible to iterate the Born series to obtain higher-order approximations of the Green function. But it is neither easy to find an analytical formula nor it makes significant contributions. So, the leading order Born-Oppenheimer approximation suffices. We focus on $\beta_\kappa$ under this approximation, which gives
\begin{eqnarray}
\beta_\kappa=-\frac{i}{2\kappa}\int_{-\infty}^{\infty} 
e^{-2i\kappa x'}V(x')dx'\,.
\end{eqnarray}

Since $V(x)\propto \calH'+\calH^2$, $V(x)$ vanishes sufficiently fast ($O(x^{-2})$ or faster) in the limit $x\to \pm\infty$. Thus the integral $\int_{-\infty}^\infty dx e^{-2i\kappa x}V(x)$, which gives $\beta_\kappa$, converges uniformly. It is then reasonable to assume that, for sufficiently large $\kappa$, the integrand vanishes exponentially in the limit $|x|\to\infty$ in the lower half complex $x$-plane. It then follows that the integral can be deformed to a contour integral by adding a segment of an infinitely large semi-circle on the lower half complex plane, namely $x=re^{i\phi}$ where $r\to\infty$ and $0>\phi>-\pi$.  The definition of the energy spectrum is only meaningful for subhorizon modes $\kappa\gg 1$, and the coefficients of the two solutions \eqref{homogeneous solution} inferred from \eqref{Born} become two contour integrals, which can be given by the residuals of the poles of the effective potential $V(x)$. 

Let  $\{x_{j}\}$ be the poles of order  $\{n_{j}\}$ of $V(x)$ in the lower half complex $x$-plane. Using the residue theorem, we can get the leading part of coefficient,
\begin{align}
\label{beta_residue}
\beta_\kappa = \frac{\pi}{\kappa} \sum_{j}\lim_{x\to x_j}\left[\frac{(e^{-2i\kappa x}V(x)(x-x_{j})^{n_{j}})^{[n_{j}-1]}}{(n_{j}-1)!}\right]\,.
\end{align}

The UV limit can help us remove most of the terms in the above residuals, by 
 noting that $e^{-2i\kappa x_i}$ is the only factor that depends on $\kappa$.
Hence, we may retain only the residue which is closest to the real axis (i.e., the smallest imaginary part in magnitude), which we denote by $x_p$. 
The Bogoliubov coefficient $\beta_\kappa$ is then given approximately by
\begin{equation}\label{profile}
	\beta_\kappa\approx\lim_{x\to x_p}\left[(e^{-2i\kappa x}V(x)(x-x_{p})^{n_{p}})^{[n_{p}-1]}\right]\,.
\end{equation}
Applying the Leibniz rule, we have
\begin{align}
    \left(e^{-2i\kappa x} f(x)\right)^{[n]}=\sum_{i=0}^n C_n^i (-2i \kappa)^{i} e^{-2i\kappa x}f(x)^{[n-i]}.
\end{align}
Therefore the leading term of \eqref{profile} is
\be
    \beta_\kappa\propto \kappa^{(n_p-1)}\left(1+O(\kappa^{-1})\right) e^{2\ImPart{[x_p]}\kappa}. 
\ee
In other words, $\beta_\kappa$ consists of the prefactor determined by the order of the dominant pole and the exponential decay rate determined by the imaginary part of the pole.

\section{Coefficients calculation for specific case}
\label{coefficients}
In order to calculate the gravitons produced $\beta_\kappa$ during this transition from inflation to radiation- or $w$-dominated era, we first need to find the solution of the transcendental equation, 
\begin{eqnarray}
\frac{1}{\calH}=\frac{1}{2\calH_t}\left[\Big((1+W)\tanh{(x/{\Delta x})}-(1-W)\Big)x+2\right]=0\,,
\end{eqnarray} where $W=({3w+1})/{2}$ and $W=0$ for the radiation case. 
Introducing $z=i x/{\Delta x}$, one sees that there exists a solution $z_p$ that satisfies $(1+W)\tan z_p={2}/({z_p {\Delta x}})+i(1-W)$ in the region $0<\mathrm{Re}[z]<\pi/2$. 
Expanding the denominator of $\calH$ in the neighborhood of $x_p=-i z_p{\Delta x}$, we have
\begin{align}
   \frac{1}{\calH}=&\frac{1}{2\calH_t}
   \left(\frac{1+W}{\Delta x}\frac{x_p}{\cosh^2(x/\Delta x)}
   -\frac{2}{x_p}\right)(x-x_p)
   +\mathcal{O}\left((x-x_p)^2\right)\notag\\
    =&-\frac{i}{2\calH_t}\left(\frac{2}{z_p{\Delta x}}
    +\frac{z_p(1+W)}{\cos^2z_p}\right)(x-x_0)+\mathcal{O}(x-x_0)^2.
\end{align}
Thus the conformal Hubble is expanded as
\begin{align}
    \label{cHexpand}
     \cH(x) = \cH_t\frac{i}{R(z_p)(x-x_p)}+\mathcal{O}(1)\,;\quad
     R(z)\equiv\frac{1}{z{\Delta x}}+\frac{z(1+W)}{2\cos^2z}\,.
\end{align}
Meanwhile, the coefficient $\beta_\kappa$ only depends on the residue of $V(x)$ at point $x_p$, then we can write $V(x)$ into the sum of $\cH'$ and $\cH^2$ and integrated them separately,
\begin{align}
    &\beta_\kappa=\calH_t^{-2} \int (ik \cH + \cH^2) e^{-2i\kappa x} d x.\notag\\
\end{align}
The residue of $V(x)$ is obvious with the Laurent expansion
\begin{align}
     &\lim_{x\to x_p}\cH(x)\, (x-x_p)= i\frac{\calH_t}{R(z_p)},\notag\\
    &\lim_{x\to x_p}\cH^2(x)\, (x-x_p)^2= -\frac{\calH_t^2}{R(z_p)^2},\notag\\
\end{align}
Substituting these expansions into \eqref{beta_residue}, we have
\begin{align}
    \beta_\kappa&=-\frac{\pi}{\kappa} \frac{2\kappa R(z_p)-2i\kappa}{R(z_p)^2}=-\frac{2\pi (R(z_p)-2i)}{R(z_p)^2}.
\end{align}
For different kinds of post-inflation model, the only thing needed to be changed is $W$, which is used in the solution of $z_p$ and the definition of $R(z)$.

\section{The amplitude of the peak for different EoS of the post-inflation stage}\label{IRtail}
The amplitude $A$ in \eqref{fitform} can be deduced by examining the IR tail of a peak ($k\ll \mathcal{H}_t^{-1}$), which remains superhorizon during the transition period. By setting the initial and final points $1/k\gg |\eta_\mathrm{i}|,|\eta_\mathrm{f}| \gg \eta_t$, the Bogoliubov coefficients after the transition can be inferred from the conservation of superhorizon tensor perturbations,
\begin{align}\label{conserve}
    t_k(\eta_i)&= \frac{H_{\inf}}{\sqrt{2k^3}}=t_k(\eta_f)=\alpha_k u_k(\eta_f)+\beta_k u_k^*(\eta_f),\\
    t'_k(\eta_f)&=\alpha_k u'_k(\eta_f)+\beta_k {u_k^*}'(\eta_f)=0,
\end{align}
where $t_k$ and $u_k$ are defined in \eqref{hankel}. For the $w$-stage, the positive frequency function can be written as 
\begin{align}
    u_k(\eta_f)&=\frac{1}{2 a(\eta_f)}\sqrt{\pi\eta} H^{(2)}_n(y)=\frac{i\Gamma{(n)}2^{n-1}}{\sqrt{\pi}a(\eta_f)}k^{-n} \eta_f^{-n+1/2}=-u_k^*(\eta_f) ~~\text{for}~~ k\eta_f\ll 1\,,
\end{align}
where $n=\left|3(1-w)/(2(1+3w))\right|$ and the second equality is because the Hankel functions $H^{(2)}_\nu(y)$ exhibit the asymptotic form,
\begin{equation}\label{eq: nneq0}
    H^{(2)}_\nu(x) = \frac{i\Gamma{(\nu)}}{\pi} \left(\frac{x}{2}\right)^{-\nu} ~~\text{for}~ \nu>0~\text{and}~ x\ll 1.
\end{equation}

With the conservation condition \eqref{conserve} and the normalization condition \eqref{normalization condition}, the Bogoliubov coefficient $\beta_k$ after the adiabatic vacuum passes through the transition is read as
\begin{equation}
    \beta_k= u_k'(\eta_f) t_k(\eta_i)a(\eta_f)^2=-\frac{in\Gamma{(n)}2^{n}a(\eta_f)}{\sqrt{2\pi}}k^{-n-3/2} \eta_f^{-n-1/2} H_{\inf},
    \label{betakforw}
\end{equation}
where $u'_k=-2n u_k/\eta$ since $a\propto \eta^{n+1/2}$ for $1>w>-1/3$. For the model with the $w$-stage after inflation, we have
\begin{equation}\label{scale_fixed}
    a(\eta_f)\eta_f^{-n-1/2}= a_\rh \left(a_\rh H_\rh W\right)^{n+1/2},
\end{equation}
where $W=(1+3w)/2$ and the suffix `rh' stands for the quantities at the end of the reheating stage. Therefore, \eqref{betakforw} can be simplified as
\begin{equation}\label{IRbeta}
    \beta_k= -\frac{i\Gamma{(n+1)}}{4\sqrt{\pi}W}\left(\frac{k}{2 a_\rh H_\rh W}\right)^{-n-3/2}\ \frac{H_{\inf}}{H_\rh}.
\end{equation}
Substituting the above \eqref{IRbeta} into \eqref{Ogw}, the spectrum of the IR tail of a peak is given by:
\begin{align}\label{IROgw}
    \Omega_{\GW,0}(f) &= \frac{\Gamma(n+1)^2}{12\pi^3\mpl^2 } 
    \left({2W}\right)^{2}
    \left(\frac{k}{2W a_\rh H_\rh}\right)^{-2n+1}  \frac{a_\rh^4 H_{\inf}^2}{a_0^4 H_0^2}H_\rh^2
    \nonumber\\
    &=\Omega_r \frac{g(T_\rh)}{g(T_0)}\frac{ \Gamma(n+1)^2}{12\pi^3\mpl^2}\left(\frac{f}{f_\rh}\right)^{-2n+1}H_{\inf}^2 \left(2W\right)^{2n+1}, 
\end{align}
where $f_{\rh}= {a_\rh H_\rh}/({2\pi a_0})$. For the inflation-radiation model, we can similarly express the scale factor dependence in \eqref{scale_fixed} in terms of the quantities at the matter-radiation equality, $a_\text{eq}$ and $H_\text{eq}$.

We can fix the normalization of the scale factor by setting the current value to unity, $a_0=1$.
Then the frequency of the peak is related to the conformal Hubble parameter at the transition as $f_\text{p}=\calH_t/(2\pi a_0)=\calH_t/(2\pi)$.
Note that the maximal conformal Hubble parameter $\calH_t$ depends on the width $\Delta x = \Delta \eta\,\calH_t$ of our parameterization \eqref{stiff Hubble}.
This implies that the peak value of the conformal Hubble parameter for the finite width case is slightly different from that for the instantaneous transition. Let us set
\begin{equation}
	\calH_t(w,\Delta x)=\frac{\calH_t^{\text{inst}}}{F(w,\Delta x)}\,,
\end{equation}
where $F(w,0)=1$ and $\calH_t^{\text{inst}}$ is the maximal conformal Hubble parameter in the instantaneous case,
\begin{equation}
	\calH_t^{\text{inst}}= \left(a_\rh H_{\inf}\left( a_\rh H_\rh\right)^{n+1/2}\right)^{1/(n+3/2)}.
\end{equation}

Substituting the peak frequency $f^{\text{inst}}_\text{p}=\calH^{\text{inst}}_t/(2\pi)$ into \eqref{IRbeta} and \eqref{IROgw}, the Bogoliubov coefficient of the peak mode and the peak amplitude in the instantaneous case are given by
\begin{align}
    \beta^{\text{inst}}_{\text{p}}&= -\frac{i\Gamma{(n+1)}}{2\sqrt{\pi}}\left({2W}\right)^{n+1/2},\\
    A^{\text{inst}}&=\Omega_{\GW,0}(f_\text{p})=\frac{ \left(2\pi f^{\text{inst}}_p\right)^4 ~\Gamma{(n+1)}^2 }{12\pi^3\mpl^2 H_0^2} \left({2}W\right)^{2n+1}.
\end{align}
These quantities are related to those for the smooth transition model as
\begin{equation}
    \beta_{\text{p}}= \beta^{\text{inst}}_{\text{p}} F(w,\Delta x)^{n+3/2},\quad A= A^{\text{inst}} F(w,\Delta)^{2n+3}.
\end{equation}
A larger width implies a smoother transition, meaning a smaller conformal Hubble parameter at the peak. For the post-inflationary stage we are interested in with $0 < w < 1$ ($n + 3/2 > 0$), this means $F(w,\Delta)>1$ for $\Delta>0$, and it is monotonically increasing with $\Delta$.
Thus the coefficient of the peak amplitude-frequency relation $A\propto f_\text{p}^4$ will be larger. Nevertheless, numerical results indicate that the factor,
\begin{equation}
    G(w,\Delta)\equiv F(w,\Delta)^{2n+3}\,,
\end{equation}
is only $\mathcal{O}(1)$. For example, $G(1/3,1)=1.7$, $G(1,1)=2.0$, and $G(1/3,2)=3.9$.

Note that although the asymptotic forms of the Hankel functions are different when $n=0$,
\begin{equation}
    H^{(2)}_0(x) =-i \frac{2}{\pi}\ln{x}\,,
\end{equation}
the Bogoliubov coefficient is obtained as
\begin{align}
    \beta^{n=0}_k=  -i \frac{H_{\inf}}{\sqrt{2\pi }}k^{-3/2}\eta_f^{-1/2} a(\eta_f)=-i \frac{H_{\inf}}{\sqrt{\pi }}k^{-3/2}\sqrt{a_\rh^3 H_\rh},
\end{align}
which is exactly equal to that of \eqref{IRbeta} in the limit $n\to0$. 
Thus, all the formulas for $n>0$ are valid in the limit $n\to0$.

\bibliography{2023.bib}
\bibliographystyle{apsrev4-1}

\end{document}